\documentclass[a4paper,11pt]{article} 
\usepackage{bm}
\usepackage[T2A]{fontenc}			
\usepackage[utf8]{inputenc}			
\usepackage[english]{babel}	
\usepackage{graphicx}
\graphicspath{}
\DeclareGraphicsExtensions{.pdf,.png,.jpg}
\usepackage{amsmath,amsfonts,amssymb,amsthm,mathtools,verbatim,alltt} 
\usepackage{braket}
\usepackage{wasysym}
\usepackage{cancel}
\usepackage{setspace}
\usepackage{slashed}
\usepackage{MnSymbol}
\usepackage{cite} 
\usepackage[margin=1.5cm]{geometry}
\usepackage{authblk}
\everymath{\displaystyle}
\usepackage{mathrsfs}

\usepackage[colorlinks = true,
linkcolor = blue,
urlcolor  = blue,
citecolor = blue,
anchorcolor = blue]{hyperref}

\date{}
\title{Entanglement viscosity to entropy density ratio for spin-3/2 theory}
\author[,1,2,3]{R. V. Khakimov\thanks{\texttt{khakimovrv@my.msu.ru}}}
\author[,1,2]{G. Yu. Prokhorov\thanks{\texttt{prokhorov@theor.jinr.ru}}}
\author[,1,2,3]{O. V. Teryaev\thanks{\texttt{teryaev@jinr.ru}}}
\affil[1]{Joint Institute for Nuclear Research, Joliot-Curie str. 6, Dubna 141980, Russia}
\affil[2]{NRC Kurchatov Institute, Moscow, Russia}
\affil[3]{Physics Department, Lomonosov Moscow State University, 1-2 Leninskie Gory, Moscow 119991, Russia}

\begin{document}
\maketitle
\begin{abstract}
It is known that the Minkowski vacuum appears as a thermal medium to an accelerated observer due to the renowned Unruh effect. More recently, it has been shown that at least for lower-spin fields this medium also exhibits a non-zero ``entanglement'' shear viscosity, which saturates the fundamental Kovtun–Son–Starinets (KSS) bound, $\eta/s = 1/4\pi$. We test the universality of this result for higher spins by computing the entanglement viscosity for spin-3/2 fields within the Rarita–Schwinger–Adler (RSA) theory. Strikingly, we obtain a negative viscosity. However, computing the entropy density using the modular Hamiltonian expansion method, we find it is also negative, and the viscosity to entropy ratio saturates the KSS bound. To clarify the origin of the negativity, we use another approach of Zubarev density operator, which gives positive entropy. 
We also show that RSA theory has many features of a conformal field theory.
\end{abstract}

\section{Introduction}
There is a deep relationship between hydrodynamics, quantum field theory, and gravity. A striking example is the celebrated Kovtun–Son–Starinets (KSS) bound \cite{Policastro:2001yc, Kovtun:2004de}
\begin{eqnarray}
    \frac{\eta}{s} \ge \frac{1}{4\pi},
    \label{KSS}
\end{eqnarray}
which states the existence of the minimal value for the viscosity $\eta$ to the entropy density $s$ ratio. Originally the bound (\ref{KSS}) was obtained from the consideration of the absorption of low-energy gravitons by a black brane within the framework of the holographic approach. However it turns out to hold in a surprisingly wide range of systems \cite{Chiofalo:2025jph, Schafer:2009dj, Parikh:1997ma, Harris:2023tti}. 
\\\\
One of the intriguing examples where the KSS bound  (\ref{KSS}) appears outside holography is the thermal radiation in an accelerated frame \cite{Chirco:2010xx}.
It is well-known that, since in this case quantum fields are limited by a Rindler horizon, the quantum vacuum is perceived as a thermal state, with the so-called Unruh temperature
\begin{eqnarray}
    T_U=\frac{a}{2 \pi},
    \label{unruhtemp}
\end{eqnarray}
that is a famous Unruh effect \cite{Unruh:1976db, Bisognano:1975ih, Bisognano:1976za, Crispino:2007eb}. Here $a=\sqrt{-a_{\mu}a^{\mu}}$ is the modulus of the 4-acceleration of the reference frame. Surprisingly, it was shown  that Unruh radiation is also characterized by finite shear viscosity. Given that the very origin of thermal radiation in this case is deeply related to quantum entanglement, the corresponding viscosity was called the entanglement viscosity. 
\\\\
Initially, it was derived for scalar field \cite{Chirco:2010xx}, and then for Dirac and Maxwell fields \cite{Lapygin:2025zhn} and recently in the general case of arbitrary conformal and nonconformal theories in \cite{Prokhorov:2026swu}. Moreover, in \cite{Prokhorov:2026swu}, universal formulas for the entanglement viscosity were obtained in terms of fundamental spectral functions. This made it possible to demonstrate the relationship between renormalization-group flows and thermodynamic irreversibility, both arising, in this context, from unitarity.
\\\\
A significant result of \cite{Chirco:2010xx, Lapygin:2025zhn, Prokhorov:2026swu} was the demonstration that for free lowest spin fields, and in general for any conformal theory in four dimensions, the entanglement viscosity exactly saturates the KSS bound (\ref{KSS}) \footnote{Notably, the KSS bound is saturated in this case only for quantities averaged over the distance to the horizon, while locally the viscosity to entropy density ratio includes the speed of sound, as has been shown recently for an any (conformal and nonconformal) theory in any number of dimensions \cite{prep2}.}. 
\\\\
Despite this, there is no firm certainty that relation (\ref{KSS}) will hold for entanglement viscosity in the case of the theories of fields with spins $S > 1$, due to the well-known difficulties of higher spin theories. In our paper, we fill this gap and consider the so-called Rarita–Schwinger–Adler (RSA) theory, which describes massless Rarita-Schwinger spin-3/2 fields $\psi_{\mu}$ coupled with additional $\lambda$ spin-1/2 field \cite{Adler:2017shl}. This additional field is nonpropagating and is needed to shift the pole in the Dirac bracket in the weak electromagnetic field limit, which is necessary for the consistency of perturbation theory.
It is of interest to check whether the universal behavior observed for entanglement viscosity of fields with lower spins \cite{Chirco:2010xx, Lapygin:2025zhn, Prokhorov:2026swu} holds in this case.
\\\\
For this purpose we compute both the shear viscosity and the entropy density for spin 3/2 fields in Rindler space. Using the Kubo formalism adapted to the Rindler spacetime \cite{Chirco:2010xx}, we evaluate the correlator with two energy-momentum tensor operators and obtain the viscosity for spin-3/2 fields. Interestingly, we find that the resulting shear viscosity is negative, in sharp contrast to the positive result obtained for lower spins \cite{Chirco:2010xx, Lapygin:2025zhn, Prokhorov:2026swu}. This formally indicates a hydrodynamic instability of Unruh radiation in the RSA model.
\\\\
On the next step, we compute the entropy density, using method, based on the perturbation theory with a modular Hamiltonian \cite{Rosenhaus:2014woa, Smolkin:2014hba, prep2}. Finite temperature and acceleration effects can be described by considering fields in Euclidean Rindler space \cite{Dowker:1994fi}, which is a space with a conical singularity. Then, the entropy, as the temperature derivative of the pressure, can be obtained as a linear term in the expansion of the energy-momentum tensor in powers of a small angular deficit. This linear term is described by the correlator of two energy-momentum tensors, according to  \cite{Smolkin:2014hba}.  Interestingly, entropy density within this approach also turns out to be negative and moreover, the corresponding ratio with the negative viscosity saturates (\ref{KSS}). Thus we demonstrate the universality of (\ref{KSS}) also for higher spins.
\\\\
However, the negativity of viscosity and entropy looks very unusual. To clarify this, we then apply another method to find entropy, based on the equilibrium form of Zubarev density operator, which includes acceleration via an effective macroscopic interaction term with the boost operator and does not require the direct use of Rindler coordinates \cite{Becattini:2023ouz, Buzzegoli:2017cqy, Becattini:2017ljh, Lapygin:2025zhn}. 
Both approaches yield the same answer for the case of lower spins \cite{prep2}. However, for the RSA theory, the second method with Zubarev operator leads to positive entropy. Moreover, within this approach, the energy-momentum tensor and entropy density of a massless field with spin 3/2 at finite temperature and acceleration corresponds (assuming additivity of the contributions of fields with different spins) exactly to spin 1/2, possibly indicating a kind of ``spin universality''.
\\\\
Moreover, while in the first method, with expansion in powers of the modular Hamiltonian, the correlator is calculated exactly at $T=T_U$, that is, ``on-shell'' (no conical singularity), then the Zubarev density operator approach is ``off-shell'', since the limit $T \to T_U$ is considered. Thus, the ``on-shell'' and ``off-shell'' calculations give different results. We speculate on the connection between this discrepancy and the similar discrepancy for the Schwinger-DeWitt coefficients discussed in \cite{Fursaev:1996uz}. We also show an analogy with the problems when considering spin-3/2 fields in an external electromagnetic field, discussed in \cite{Adler:2017shl}.
\\\\
We also show that using the example of entanglement viscosity, we can talk about the so-called ``wandering'' Planck constant - if, for example, in the famous membrane approach \cite{Parikh:1997ma}, Planck constant $\hbar$ in the KSS bound arises from the quantum entropy of a black hole, while the viscosity is purely classic, then in the case of entanglement viscosity, the situation is reversed and quantumness comes precisely from viscosity. We also discuss a similar property in the context of angular momentum quantization and the Kolmogorov cascade.
\\\\
The paper is organized as follows. Section \ref{framework} briefly describes the Rarita-Schwinger-Adler theory and the methods for calculating the viscosity and entropy density of quantum fields in Rindler spacetime. Section \ref{RSAVISCOSITY} presents the calculation of the shear viscosity in the RSA theory using the Kubo formalism, and it is shown that the viscosity is negative. In Section \ref{entropy1}, we calculate the entropy density using the modular Hamiltonian expansion and show that the entropy is also negative, and viscosity to entropy density ratio saturates the KSS bound. In Section \ref{entropy2}, we use an off-shell method with the Zubarev operator to the find energy-momentum tensor at finite temperature and acceleration, and find that the entropy in this approach is positive.
In Section \ref{discussion}, we discuss the meaning and consequences of the obtained results - in particular, the significance of the negative viscosity, the discrepancies between different methods of calculating the energy-momentum tensor and entropy, the possible universality of the energy-momentum tensor for various spins, as well as the difference in the origin of the Planck constant in the KSS bound for different systems, where it is saturated.
\\\\
We use the notations and conventions $ \delta_{\mu\nu} = {\text{diag}(1,1,1,1)} $, $ \eta_{\mu\nu} = {\text{diag}(1,-1,-1,-1)} $, $\varepsilon^{0123}=1$, 
$ \gamma_5=i \gamma^0\gamma^1\gamma^2\gamma^3$, and the system of units $ e=\hbar=c=k_B=1 $. The summation is assumed over repeated indices, regardless of their location, i.e. $P_{\mu}P_{\mu}= P_{0} P_{0} + \bm{P}\cdot \bm{P} $ implies contraction with Euclidean metric, and $ P_{\mu}X^{\mu}= P_{0} X_{0} - \bm{P}\cdot \bm{X} $ includes summation with a non-Euclidean metric.

\section{Theoretical framework}
\label{framework}

\subsection{Rarita-Schwinger-Adler model}

The Rarita–Schwinger theory provides the standard framework for describing fields with spin-3/2 and plays a central role in supergravity \cite{Freedman:2012zz} and grand unification models \cite{Adler:2014pga}, contributing to the elimination of anomalies there. It has also found applications for describing condensed matter systems such as Rarita-Schwinger-Weyl semimetals \cite{Boettcher:2019dtq}. However, the traditional Rarita-Schwinger theory has a number of well-known problems if used outside of supergravity \cite{Velo:1969bt, Adler:2015yha, Adler:2017shl, Adler:2019zxx}, for example, the  singularity of the Dirac bracket in the weak electromagnetic field limit, the occurrence of superluminal modes, and the presence of unphysical degrees of freedom. Some of these problems can be solved within the so-called Rarita-Schwinger-Adler model developed in \cite{Adler:2017shl}, which includes an additional spin-1/2 field $\lambda$ coupled to the Rarita-Schwinger field $\psi_{\mu}$. In flat Minkowski spacetime, the corresponding Lagrangian has the form
\begin{eqnarray}
\mathcal{L} = - \varepsilon^{\alpha \beta \mu \nu} \Bar{\psi}_{\alpha} \gamma_5 \gamma_{\mu} \partial_{\nu} \psi_{\beta} + i \Bar{\lambda} \gamma^{\mu} \partial_{\mu} \lambda - im \Bar{\lambda} \gamma^{\mu} \psi_{\mu} + im \Bar{\psi}_{\mu} \gamma^{\mu} \lambda\,,
\label{theory}
\end{eqnarray}
where the parameter $m$ can be called ``coupling mass'', and following \cite{Adler:2017shl}, we will consider the limit $m \to \infty$. An additional field $\lambda$ with spin-1/2 was added to the theory to solve the problem with the singularity in the Dirac bracket in the limit of a weak electromagnetic field and the jump in the number of degrees of freedom when moving from free Rarita-Schwinger fields to those interacting with the electromagnetic field. Note that the propagator of this additional field vanishes $\langle \lambda\bar{\lambda}\rangle=0$, meaning it is non-propagating.
\\\\
The energy-momentum tensor can be constructed from the Lagrangian by varying the background metric. Thus, energy-momentum tensor for RSA theory has the form \cite{Prokhorov:2022rna}
\begin{eqnarray} \label{EMT}
T^{\mu\nu} &=&
\frac{1}{2}\,\epsilon^{\lambda\nu\beta\rho}\,\bar{\psi}_\lambda \gamma^5 \gamma^\mu \partial_\beta \psi_\rho
+ \frac{1}{8}\,\partial_\eta \!\left( \epsilon^{\lambda\alpha\nu\rho}\,\bar{\psi}_\lambda \gamma^5 \gamma_\alpha [\gamma^\eta, \gamma^\mu] \psi_\rho \right)+ \\ \nonumber
&&+ \frac{i}{4}\left(\bar{\lambda} \gamma^\nu \partial^\mu \lambda - \partial^\mu \bar{\lambda} \gamma^\nu \lambda \right)
+ \frac{i}{2} m \left(\bar{\psi}^\mu \gamma^\nu \lambda - \bar{\lambda} \gamma^\nu \psi^\mu \right)
+ (\mu \leftrightarrow \nu)\, .
\end{eqnarray}
It can be seen from the expression (\ref{EMT}) that the energy-momentum tensor is traceless
\begin{eqnarray}
    T^{\mu}_{\mu}=0\, .
\end{eqnarray}
This means that the RSA theory is scale-invariant, indicating that it may also be conformally invariant, which we will discuss in more detail later. The propagators can be obtained in the usual way by inverting the Lagrangian and in the coordinate representation have the form \cite{Prokhorov:2022rna}
\begin{equation}
    \begin{cases}
        G^{\psi \bar{\psi}}_{\mu \nu} (x) = \langle 0| \psi_{\mu}(x) \bar{\psi}_{\nu}(0) |0\rangle_M = \frac{i}{4 \pi^2 \bar{x}^4} \bigg( \gamma_{\nu} \slashed x \gamma_{\mu} -2\left(1+\frac{4}{m^2\bar{x}^2}\right) \big( \eta_{\mu \nu} \, \slashed x + \gamma_{\mu} x_{\nu} + \gamma_{\nu} x_{\mu} \big) \\
\qquad\qquad\qquad\qquad\qquad\qquad\quad\,\,\,\,\, + 8 \left(1+\frac{6}{m^2\bar{x}^2}\right) \frac{x_{\mu} x_{\nu} \slashed x}{\bar{x}^2} \bigg)\,,\vspace{0.2 cm} \\
        G^{\lambda \bar{\psi}}_{\mu} (x) = \langle0| \lambda(x) \bar{\psi}_{\mu}(0) |0\rangle_M =  \frac{i}{2 \pi^2 m \, \bar{x}^4} \Big( \gamma_{\mu} - \frac{4 x_{\mu} \slashed x}{\bar{x}^2} \Big)\,, \vspace{0.2 cm} \\ 
        G^{\psi \bar{\lambda}}_{\mu} (x) = \langle 0| \psi_{\mu}(x) \bar{\lambda}(0) |0\rangle_M =  -\frac{i}{2 \pi^2 m \, \bar{x}^4} \Big( \gamma_{\mu} - \frac{4 x_{\mu} \slashed x}{\bar{x}^2} \Big)\,,\\ 
        G^{\lambda \bar{\lambda}}_{\mu} (x) = \langle 0| \lambda(x) \bar{\lambda}(0) |0\rangle_M = 0\,.
    \end{cases}
    \label{xpropagators}
\end{equation}
where $\bar{x}^2=x^2-i\varepsilon x_0$, $\varepsilon \to 0^+$,  the average $\langle 0| ... |0\rangle_M$ is taken over the Minkowski vacuum and $x_{\mu}$ is the Minkowski coordinate. The shift in the poles corresponds to the choice of a non-time-ordered propagator, which is needed within the linear response theory. Multiplying by time in $-i\varepsilon x_0$ causes both poles $x_0=\pm\sqrt{\bm{x}^2}$ to be shifted upward from the real axis.


\subsection{Rindler coordinates and stretched horizon}

The Rindler wedge represents the region of Minkowski spacetime accessible to an observer undergoing uniform acceleration. Introducing the Rindler coordinates $\mathcal{X}=(\tau,\text{x},\text{y},\xi)$  related to the Minkowski coordinates $X=(t,\text{x},\text{y},\text{z})$ as
\begin{equation}
    \begin{cases}
    t = \xi \sinh( \tau), \vspace{0.1 cm} \\ 
    z=  \xi \cosh( \tau),\\
    \text{x} = \text{x}, \\
    \text{y} = \text{y},
\end{cases}
\label{rindler}
\end{equation}
the metric takes the form
\begin{eqnarray}
    \label{metric}
    ds^2=\xi^2 d \tau^2-d\text{x}^2-d\text{y}^2-d\xi^2,
\end{eqnarray}
which covers only the quadrant $z>|t|$, that is, the right Rindler wedge. The worldlines of constant $\xi$ correspond to uniformly accelerated observers with proper acceleration $a=1/\xi$, while the hypersurface $\xi=0$ acts as a causal horizon. When a quantum field is restricted to the Rindler wedge, the Minkowski vacuum appears as a thermal state with respect to the Rindler Hamiltonian generating boosts. This is the manifestation of the Unruh effect \cite{Unruh:1976db}, according to which an observer with proper acceleration $a_{\mu}$ experiences a thermal bath at the Unruh temperature (\ref{unruhtemp}). Expectation values of operators confined to the wedge are thus equivalent to thermal averages \cite{Unruh:1983ac}
\begin{eqnarray}\label{UnruhWeiss}
    \langle 0 | \hat{O}_R(t,x^i,z)|0 \rangle_M = Tr [e^{2\pi H_R} \hat{O}_R(\tau,x^i,\xi)]\,,
\end{eqnarray}
so that the Minkowski vacuum becomes a mixed state after tracing over field degrees of freedom beyond the Rindler horizon.
\\\\We will regularize the distance to the horizon with a small parameter $l_c$, which goes back to the famous membrane paradigm \cite{Parikh:1997ma, Thorne:1986iy}. The membrane paradigm  provides a complementary perspective of the black hole, one of the essential elements of which is the replacement of the horizon by a timelike surface, the so-called stretched horizon, located slightly in front of the true horizon. In our case with the Rindler space, this surface is located at $\xi=l_c$, where $l_c>0$ is assumed to be small. This stretched horizon behaves as an effective two-dimensional membrane endowed with dissipative hydrodynamic properties.
\\\\
We are, however, not interested in the \textit{classical} membrane itself and its associated viscosity, but in the \textit{quantum} thermal radiation, Unruh radiation, that exists above this membrane at 
\begin{eqnarray}\label{lcc}
\xi>l_c\,.
\end{eqnarray}
Thus, parameter $l_c$ plays a purely regularizing role when considering volume integrals in our case. 

\subsection{Entropy in the Rindler spacetime}

To calculate the ratio (\ref{KSS}), we need to know the entanglement entropy density in Rindler spacetime. We follow the approach developed in particular in \cite{Becattini:2023ouz}, according to which the entropy density can be calculated from the heat balance equation as the derivative of pressure at constant acceleration
\begin{eqnarray}
    s_{loc} = \frac{\partial p}{ \partial T} \Bigg|_a ,
    \label{sloc}
\end{eqnarray}
where the subscript \textit{loc} indicates that the quantity corresponds to a \textit{local} point at a certain distance from the horizon. The corresponding energy density and pressure can be found from the mean value of the energy-momentum tensor operator
\begin{eqnarray}
    \varepsilon = \langle \hat{T}_{\mu\nu} \rangle u^{\mu} u^{\nu}, \quad p=-\frac{1}{3} \langle \hat{T}_{\mu\nu} \rangle P^{\mu\nu},
\end{eqnarray}
where $u^{\mu}$ is the 4-velocity of the medium, and $P^{\mu\nu}=g^{\mu\nu}-u^{\mu}u^{\nu}$ is the projector to the surface  orthogonal to $u^{\mu}$. Since we are considering the Minkowski vacuum state, though from the point of view of an accelerated observer, we can substitute the temperature value $T=T_U=a/2\pi$ in (\ref{sloc}) after calculating the derivative \cite{Becattini:2017ljh}. If we then take into account that $a = 1/\xi$, according to (\ref{rindler}), then the entropy density will be a function only of the distance to the horizon
\begin{eqnarray}
\label{rule}
    s_{loc} (T=T_U,a) \xrightarrow[a \to 1/\xi] {} s_{loc} (\xi),
\end{eqnarray}
We are also interested in the total entropy per unit area of the Rindler horizon and therefore we will integrate it over
$\xi$ from $l_c$ to infinity 
\begin{eqnarray}
\label{intrule}
    s= \int_{l_c}^{\infty} d \xi \, s_{loc}(\xi)
\end{eqnarray}
This entropy can be called \textit{global}.

\subsection{Linear response theory in the right Rindler wedge}

As stated above, it turned out that Unruh radiation itself has viscosity \cite{Chirco:2010xx}. From the linear response theory, we can write a Kubo formula to calculate the shear viscosity \footnote{Kubo formula can be written in slightly different forms, using the relationship between different Green functions \cite{Laine:2016hma, Son:2002sd}. 
In particular, the viscosity we are considering can be obtained from a non-time-ordered Green function, that is the Wightman function, see \cite{Chirco:2010xx}, or directly from the retarded one \cite{Prokhorov:2026swu}, and it can be shown that both methods are equivalent (for example, different Kubo formulas may differ by the presence of a factor $1/\omega$, which, however, does not affect the result).}
\begin{eqnarray}
    \eta_{\,loc}(\xi') = \pi \, \lim_{\omega \to 0^+} 
\int_{\ell_c}^{\infty} d\xi  
\iint_{-\infty}^{\infty} d\text{x} \,d\text{y} \int_{-\infty}^{\infty} d\tau \, 
e^{i \omega \tau} \,
 \, \xi \, \xi' \;
\langle 0| \hat{T}_{\text{x}\text{y}}(\tau, \text{x}, \text{y}, \xi) \, \hat{T}_{\text{x}\text{y}}(0,0,0, \xi') |0 \rangle_M\,.
\label{etakubo}
\end{eqnarray}
Signature $loc$ means that the viscosity refers to a specific point in space, at a distance $\xi'$ from the horizon. We will call this viscosity, by analogy with entropy (\ref{rule}), \textit{local} viscosity. Following (\ref{lcc}), we regularize the integral over $\xi$ from below by the ``membrane thickness'' $l_c$. However, we can also consider the total viscosity per unit area of the horizon 
\begin{eqnarray}
    \eta = \int_{l_c}^{\infty} d \xi \, \eta_{\,loc}(\xi)\,,
\label{etakubo1}
\end{eqnarray}
which we will call \textit{global} viscosity. It should be noted that (\ref{etakubo}) corresponds only to the point $T=T_U$, that is, the Minkowski vacuum state.
\\\\
In fact, in (\ref{etakubo}) we apply the property (\ref{UnruhWeiss}) from \cite{Unruh:1983ac} that for the Minkowski vacuum state, thermal average in Rindler space can be obtained from the Minkowski vacuum correlator in Minkowski space. Thus, effectively, we can find the usual vacuum correlator in usual Minkowski coordinates, and then move to Rindler coordinates only before taking the Fourier transform.
This significantly simplifies the calculations, and keeping this in mind, let us go directly to the derivation of viscosity for the Rarita-Schwinger-Adler theory in the Rindler spacetime.

\section{Viscosity in the Rarita-Schwinger-Adler model}
\label{RSAVISCOSITY}

Now, let us take a closer look at the details of the viscosity calculation for the RSA theory (\ref{theory}).  When calculating, we will rely on the formula (\ref{etakubo}). As can be seen from (\ref{EMT}) energy-momentum tensor can be divided into four parts
\begin{eqnarray}
     \hat{T}^{\mu\nu} = \hat{T}^{\mu\nu}_{\bar{\psi} \psi} + \hat{T}^{\mu\nu}_{\bar{\psi} \lambda} + \hat{T}^{\mu\nu}_{\bar{\lambda} \psi} + \hat{T}^{\mu\nu}_{\bar{\lambda} \lambda}.
     \label{SETIK}
\end{eqnarray}
The two-point correlator therefore has the form
\begin{eqnarray}
\label{terms2corr}
    \langle \hat{T}^{\mu\nu} \, \hat{T}^{\alpha\beta} \rangle =  \langle \hat{T}^{\mu\nu}_{\bar{\psi} \psi} \hat{T}^{\alpha\beta}_{\bar{\psi} \psi} \rangle + \langle \hat{T}^{\mu\nu}_{\bar{\psi} \psi} \hat{T}^{\alpha\beta}_{\bar{\psi} \lambda} \rangle +\langle \hat{T}^{\mu\nu}_{\bar{\psi} \psi} \hat{T}^{\alpha\beta}_{\bar{\lambda} \psi} \rangle + \langle \hat{T}^{\mu\nu}_{\bar{\psi} \lambda} \hat{T}^{\alpha\beta}_{\bar{\psi} \psi} \rangle + \langle \hat{T}^{\mu\nu}_{\bar{\psi} \lambda} \hat{T}^{\alpha\beta}_{\bar{\psi} \lambda} \rangle + \langle \hat{T}^{\mu\nu}_{\bar{\lambda} \psi} \hat{T}^{\alpha\beta}_{\bar{\psi} \psi} \rangle + \langle \hat{T}^{\mu\nu}_{\bar{\lambda} \psi} \hat{T}^{\alpha\beta}_{\bar{\lambda} \psi} \rangle+...\, .
\end{eqnarray}
We will first consider the correlator in the general case, and then move on to the indices $\mu =\alpha= 1$ and $\nu =\beta = 2$ in accordance with the formula (\ref{etakubo}). All correlators (not written out explicitly in (\ref{terms2corr})) containing the tensor $\hat{T}_{\bar{\lambda} \lambda}$ as well as the terms $ \langle \hat{T}^{\mu\nu}_{\bar{\psi} \lambda} \hat{T}^{\alpha\beta}_{\bar{\lambda} \psi} \rangle$ and $ \langle \hat{T}^{\mu\nu}_{\bar{\lambda} \psi} \hat{T}^{\alpha\beta}_{\bar{\psi} \lambda} \rangle$ will give zero in the considered limit $m \rightarrow \infty$ or due to the condition of a zero $\lambda$-field propagator $\langle \bar{\lambda} \lambda \rangle = 0$.
\\\\
It is convenient to use the point-splitting procedure and represent each energy-momentum tensor as a differential operator acting on a two-point propagator. Thus, energy-momentum tensors have the form \cite{Prokhorov:2022rna}
\begin{equation}
    \begin{cases}
    \hat{T}^{\mu\nu}_{\bar{\psi} \psi} (X) = \lim_{X_1,X_2 \to X} \hat{\mathcal{D}}_{\bar{\psi} \psi}^{\mu\nu \rho \sigma}(\partial_{X_1},\partial_{X_2}) \langle \bar{\Psi}_{\rho}(X_1) \Psi_{\sigma} (X_2) \rangle, \vspace{0.2 cm} \\
    \hat{T}^{\mu\nu}_{\bar{\psi} \lambda} (X) = \lim_{X_1,X_2 \to X} \hat{\mathcal{D}}_{\bar{\psi} \lambda}^{\mu\nu \rho} \langle \bar{\Psi}_{\rho}(X_1) \lambda (X_2) \rangle, \vspace{0.2 cm} \\ \hat{T}^{\mu\nu}_{\bar{\lambda} \psi} (X) = \lim_{X_1,X_2 \to X} \hat{\mathcal{D}}_{\bar{\lambda} \psi}^{\mu\nu \rho} \langle \bar{\lambda}(X_1) \Psi_{\rho} (X_2) \rangle,
    \end{cases}
    \label{pointsplit}
\end{equation}
where $X^{\mu}=(t,\text{x},\text{y},\text{z})$,  $\partial_{X}^{\mu}=\frac{\partial}{\partial X_{\mu}}$, and operators $\hat{\mathcal{D}}$ are defined as
\begin{equation}
    \begin{cases}
    \hat{\mathcal{D}}_{\bar{\psi} \psi}^{\mu\nu \rho \sigma} = \frac{1}{2} \varepsilon^{\rho\sigma\nu\eta} \gamma_5 \gamma^{\mu} \partial^{X_2}_{\eta} - \frac{1}{8} \varepsilon^{\rho\sigma\nu\eta} \gamma_5 \gamma_{\eta} [\gamma^{\xi},\gamma^{\mu}] (\partial^{X_1}_{\xi} + \partial^{X_2}_{\xi}) + (\mu \longleftrightarrow \nu)\,,\vspace{0.2 cm} \\
    \hat{\mathcal{D}}_{\bar{\psi} \lambda}^{\mu\nu \rho} = \frac{im}{2} \Big( \gamma^{\mu} \eta^{\nu \rho} + \gamma^{\nu} \eta^{\mu \rho} \Big)\,, \vspace{0.2 cm} \\ 
    \hat{\mathcal{D}}_{\bar{\lambda} \psi}^{\mu\nu \rho} = - \frac{im}{2} \Big( \gamma^{\mu} \eta^{\nu \rho} + \gamma^{\nu} \eta^{\mu \rho} \Big)\,.
    \end{cases}
    \label{diffoperators}
\end{equation}
Now, for clarity, we present here the calculation of the first term of the expression (\ref{terms2corr}). All remaining terms can be calculated similarly. Substituting (\ref{pointsplit}) into the first term in (\ref{terms2corr}), we obtain
\begin{eqnarray}\nonumber
    \langle 0| \hat{T}^{\mu\nu}_{\bar{\psi} \psi}(X) \hat{T}^{\alpha\beta}_{\bar{\psi} \psi} (Y) |0\rangle_M &=& \lim_{X_1,X_2 \to X} \, \lim_{Y_1,Y_2 \to Y} \hat{\mathcal{D}}_{(\bar{\psi} \psi)a_1 a_2}^{\mu\nu \rho \sigma}(\partial_{X_1},\partial_{X_2})  \hat{\mathcal{D}}_{(\bar{\psi} \psi) a_3 a_4}^{\alpha\beta\delta\tau}(\partial_{Y_1},\partial_{Y_2}) \\ \nonumber
    &&\cdot\, \langle 0| \bar{\Psi}^{a_1}_{\rho}(X_1) \Psi^{a_2}_{\sigma} (X_2) \bar{\Psi}^{a_3}_{\delta}(Y_1) \Psi^{a_4}_{\tau} (Y_2) |0\rangle_M\\  
    &=&  \hat{\mathcal{D}}_{(\bar{\psi} \psi)a_1 a_2}^{\mu\nu \rho \sigma}(\partial_{X_1},\partial_{X_2})  \hat{\mathcal{D}}_{(\bar{\psi} \psi)a_3 a_4}^{\alpha\beta\delta\tau}(\partial_{Y_1},\partial_{Y_2}) G^{a_2 a_3}_{\sigma\delta} (X_2-Y_1) G^{a_4 a_1}_{\tau\rho} (X_1-Y_2)\,,
    \label{corr0}
\end{eqnarray}
where $a_i$ are bispinor indices and we have taken into account only connected contributions. Substituting the propagator from (\ref{xpropagators}) and vertex from (\ref{diffoperators}) into (\ref{corr0}), performing all differentiations directly and moving on to the limit $m\to \infty$, as a result, we obtain
\begin{eqnarray}\nonumber
    \langle 0| \hat{T}^{\mu\nu}_{\bar{\psi} \psi}(X) \hat{T}^{\alpha\beta}_{\bar{\psi} \psi} (Y) |0\rangle_M &=& \frac{1}{\pi^4}\bigg( 
    \frac{168 b^{\mu} b^{\nu} b^{\alpha} b^{\beta}}{\bar{b}^{12}}
-\frac{50 \eta^{\mu\alpha} b^{\nu} b^{\beta}}{\bar{b}^{10}}
-\frac{50\eta^{\nu\alpha} b^{\mu} b^{\beta}}{\bar{b}^{10}}
-\frac{50\eta^{\mu\beta} b^{\nu} b^{\alpha}}{\bar{b}^{10}}
-\frac{50\eta^{\nu\beta} b^{\mu} b^{\alpha}}{\bar{b}^{10}}
\\&&+\frac{8\eta^{\mu\nu} b^{\alpha} b^{\beta}}{\bar{b}^{10}}
+\frac{8\eta^{\alpha\beta} b^{\mu} b^{\nu}}{\bar{b}^{10}}
+\frac{29\eta^{\mu\alpha} \eta^{\nu \beta}}{\bar{b}^8}
+\frac{29\eta^{\mu\beta} \eta^{\nu \alpha}}{\bar{b}^8}
-\frac{33\eta^{\mu\nu} \eta^{\alpha \beta}}{2\bar{b}^8}\bigg) \,,
    \label{corrpsipsi}
\end{eqnarray}
where 
\begin{eqnarray} \label{poles}
b_{\mu} = X_{\mu}-Y_{\mu}\,, \quad \bar{b}^2 = b^2 - i \varepsilon b_0\,,
\end{eqnarray}
We emphasize that this correlator is not conformally symmetric. According to conformal symmetry, one would expect the correlator (in four dimensions) to be proportional \cite{Erdmenger:1996yc}
\begin{eqnarray}\nonumber
&&\langle 0| \hat{T}^{\mu\nu}(X) \hat{T}^{\alpha\beta}(Y)  |0\rangle_M = C_T \mathcal{I}^{\mu\nu\alpha\beta} (b)\,,\\ \nonumber
&&
\mathcal{I}_{\mu\nu\alpha\beta}(b)=\frac{I_{\mu\rho}(b)I_{\nu\sigma}(b)\mathcal{E}^{\rho\sigma}{}_{\alpha\beta}}{\bar{b}^8}\,,
\\ 
&&
I_{\mu\nu}(b)=\eta_{\mu\nu}-2\frac{b_{\mu} b_{\nu}}{\bar{b}^2}\,,\quad
\mathcal{E}_{\mu\nu\alpha\beta}=\frac{1}{2}(\eta_{\mu\alpha}\eta_{\nu\beta}+\eta_{\mu\beta}\eta_{\nu\alpha})-\frac{1}{4}\eta_{\mu\nu}\eta_{\alpha\beta}\,,
\label{II}
\end{eqnarray}
where $C_T$ is the conformal central charge. Expanding the brackets in (\ref{II}), we obtain
\begin{eqnarray}
\mathcal{I}_{\mu\nu\alpha\beta} (b) = \frac{4b_{\mu} b_{\nu} b_{\alpha} b_{\beta}}{\bar{b}^{12}}
-\frac{\eta_{\mu\alpha} b_{\nu} b_{\beta}}{\bar{b}^{10}}
-\frac{\eta_{\nu\alpha} b_{\mu} b_{\beta}}{\bar{b}^{10}}
-\frac{\eta_{\mu\beta} b_{\nu} b_{\alpha}}{\bar{b}^{10}}
-\frac{\eta_{\nu\beta} b_{\mu} b_{\alpha}}{\bar{b}^{10}}
+\frac{\eta_{\mu\alpha} \eta_{\nu \beta}}{2\bar{b}^8}
+\frac{\eta_{\mu\beta} \eta_{\nu \alpha}}{2\bar{b}^8}
-\frac{\eta_{\mu\nu} \eta_{\alpha \beta}}{4\bar{b}^8}\,.  \label{II1}
\end{eqnarray}
Comparing (\ref{corrpsipsi}) with (\ref{II1}), we see that the correlator is evidently not conformally symmetric.
\\\\
However, by calculating the other terms in (\ref{terms2corr}) in the same manner, we obtain the correlator of two total energy-momentum tensors
\begin{eqnarray}
    \langle 0| \hat{T}^{\mu\nu}(X) \hat{T}^{\alpha\beta} (Y) |0\rangle_M &=& -\frac{14}{\pi^4}  \mathcal{I}_{\mu\nu\alpha\beta} (b)\,,
    \label{corr0b}
\end{eqnarray}
that is, the full correlator turns out to be exactly conformally symmetric. However, it is characterized by a negative conformal central charge. Note that conventional Rarita-Schwinger theory does not possess conformal symmetry, and it is even more surprising that in the RSA model, the corresponding one-loop graph (in the limit $m\to \infty$) turns out to be conformally symmetric. 
\\\\
In general, the subsequent calculations turn out to be similar to lower spins \cite{Chirco:2010xx, Lapygin:2025zhn}, however, we will repeat the corresponding derivation.
Leaving only the components with $\mu=\alpha=1$ and $\nu=\beta=2$ in the expression (\ref{corr0b}) and moving to the point $X=(t,\text{x},\text{y},\text{z})$ and $Y=(0,0,0,\text{z}')$, we obtain
\begin{eqnarray}
\langle \hat{T}_{\text{xy}}(X) \hat{T}_{\text{xy}}(Y)  \rangle = \frac{-7}{\pi^4} \Bigg( \frac{8\text{x}^2\text{y}^2}{\bar{b}^{12}} + \frac{2\text{x}^2}{\bar{b}^{10}} + \frac{2\text{y}^2}{\bar{b}^{10}} + \frac{1}{\, \bar{b}^{8}}\Bigg),
    \label{G1}
\end{eqnarray}
where $\bar{b}^2$ takes the form
\begin{eqnarray}
    \bar{b}^2=t^2 -\text{x}^2 - \text{y}^2 - (\text{z}-\text{z}')^2 - i \varepsilon t.
   \label{r2}
\end{eqnarray}
Now we substitute the value for $r^2$ from (\ref{r2}) and then change the Minkowski coordinates to the Rindler ones according to (\ref{rindler}) with $z' = \xi'$
\begin{eqnarray}
    G_{\text{xy},\text{xy}}(\tau, \text{x},\text{y},\xi,\xi')=\langle \hat{T}_{xy} \hat{T}_{xy}  \rangle = \frac{-7}{\pi^4} \Bigg( \frac{8\text{x}^2\text{y}^2}{(\rho^2 + \alpha)^6} - \frac{2\text{x}^2}{(\rho^2 + \alpha)^5} - \frac{2\text{y}^2}{(\rho^2 + \alpha)^5} + \frac{1}{(\rho^2 + \alpha)^4} \Bigg),
    \label{G2}
\end{eqnarray}
where 
\begin{eqnarray}
    \alpha=\xi^2+\xi'^2-2\xi\xi'\cosh( \tau) + i \varepsilon,
\end{eqnarray}
and $\rho^2 = \text{x}^2+\text{y}^2$. For further calculation of viscosity, it is necessary to perform the Fourier transform into Rindler frequency and momentum (taking the zero-momentum limit)
\begin{eqnarray}
    \Tilde{G}_{\text{xy},\text{xy}}(\omega, \xi,\xi') = \int^{\infty}_{l_c} \xi' \, d \xi' \int^{\infty}_{l_c} \xi \, d \xi \, \int^{\infty}_{-\infty} d \tau \, e^{i \omega \tau} \iint^{\infty}_{-\infty} d\text{x} \,d\text{y} \, G_{\text{xy},\text{xy}}(\tau, \text{x}, \text{y}, \xi, \xi')
    \label{G3}
\end{eqnarray}
It is convenient to make a coordinate change to
\begin{eqnarray}
    \text{x} = \rho \cos{\theta}, \\ \nonumber
    \text{y} = \rho \sin{\theta}
\end{eqnarray}
so the coordinate integration will turn into 
\begin{eqnarray}
    \iint^{\infty}_{-\infty} d\text{x}d\text{y} \to \int^{\infty}_{0} \rho \, d \rho \int^{2\pi}_{0} d \theta.
\end{eqnarray}
Integrating over the angles and over $\rho$, we obtain
\begin{eqnarray}
    \Tilde{G}(\omega, \xi,\xi') = \frac{-14}{\pi^3}\int^{\infty}_{l_c} \xi' d \xi' \int^{\infty}_{l_c} \xi d \xi \, \int^{\infty}_{-\infty} d \tau \, e^{i \omega \tau} \int^{\infty}_0  \frac{ \alpha^2 \, \rho}{(\alpha+\rho^2)^6} d\rho = \\ \nonumber
    = -\frac{7}{5 \pi^3}\int^{\infty}_{l_c} d \xi' \int^{\infty}_{l_c} d \xi \, \int^{\infty}_{-\infty} d \tau \, e^{i \omega \tau} \frac{\xi \xi'}{(\xi^2+\xi'^2-2\xi\xi'\cosh( \tau) + i \varepsilon)^3}
\end{eqnarray}
The next step is an integration over the Rindler time $\tau$. It can be seen that the integral 
\begin{eqnarray}
    I =  \int_{-\infty}^{\infty}  \frac{e^{i \omega \tau}}{(\xi^2+\xi'^2-2\xi\xi'\cosh(\tau) + i \varepsilon)^3} d \tau
    \label{bigI}
\end{eqnarray}
contains poles of the third order at the points
\begin{eqnarray}
    \tau_0=\pm \log(\xi/\xi') +  2\pi n i, \quad \quad n=0, \pm 1, \pm 2,...
\end{eqnarray}
To calculate the integral in the complex plane, it is most convenient to close the contour through the upper half-plane with $\tau \to \tau + 2 \pi i$. Since $\cosh(\tau + 2 \pi i) = \cosh(\tau)$ and $e^{i \omega(\tau+2 \pi i)} = e^{-2\pi \omega} e^{i \omega\tau}$, we can relate the whole contour integral $I_{full}$ with the integral (\ref{bigI}) in which we are interested (more detailed description of the procedure with the image of the integration contour is given in the work \cite{Lapygin:2025zhn}).
\begin{eqnarray}
    I_{full} = I-I e^{-2 \pi \omega} = 2 \pi i \sum_{\tau = \tau_0} Res \frac{e^{i \omega \tau}}{(\xi^2 + \xi'^2 -2 \xi \xi' \cosh( \tau))^3}
\end{eqnarray}
Thus, the desired integral (\ref{bigI}) can be calculated by finding the residue at the third-order pole
\begin{eqnarray}
    I = \frac{2 \pi i}{1-e^{-2 \pi \omega}}  \, \, \frac{1}{2} \sum_{\tau=\tau_0} \lim_{\tau \to \tau_0} \frac{d^2}{d \tau^2} \Bigg[ \frac{e^{i \omega \tau}(\tau - \tau_0)}{(\xi^2+\xi'^2-2\xi\xi'\cosh( \tau) + i \varepsilon)^3} \Bigg]
\end{eqnarray}
Computing the residues and taking the limit $\omega \to 0$, we find
\begin{eqnarray}
    \Tilde{G}(0, \xi, \xi') =-\frac{7}{5 \pi^3} \int_{l_c}^{\infty}d \xi \, \xi\int_{l_c}^{\infty} d\xi' \, \xi'\frac{-3(\xi^4-\xi'^4)+2(\xi^4+4\xi^2 \xi'^2+\xi'^4)\log(\xi/\xi')}{(\xi^2-\xi'^2)^5}
\end{eqnarray}
Eventually, after performing necessary integrations and multiplying by $\pi$ the result as in (\ref{etakubo}), we obtain the final expression for viscosity
\begin{eqnarray}
    \label{viscosity}
    \eta = - \frac{7}{240 \, \pi^2 \, l_c^2}
\end{eqnarray}
which, unexpectedly, turns out to be negative.
\\\\
Although we cannot strictly isolate the contribution of the massless spin-3/2 field, since the field $\psi_{\mu}$ contains extra degrees of freedom and is coupled to $\lambda$, it is interesting to consider the contribution to the final negative viscosity from various contributions in (\ref{terms2corr}). Considering the contribution from $\psi$ only, we obtain a positive result
\begin{eqnarray}
    \eta_{\psi} =  \frac{31}{240 \, \pi^2 \, l_c^2}.
\end{eqnarray}
Meanwhile all the other correlators containing additional spin-1/2 $\lambda$ fields that will make a decisive negative contribution to the overall viscosity value
\begin{eqnarray}
    \eta_{\lambda} = - \frac{38}{240 \, \pi^2 \, l_c^2}.
\end{eqnarray}
Such as $\eta = \eta_{\psi}+\eta_{\lambda}$.

\section{Entropy from two-point correlation function: negative entropy}
\label{entropy1}

Since the vacuum of an accelerated reference frame corresponds to a thermal state with the Unruh temperature \cite{Unruh:1976db, Fulling:1972md}, there are at least two ways to calculate the entropy of such a system. The correlators of the system can be considered immediately at the Unruh temperature $T=T_U = a/ 2 \pi$, which corresponds to the calculation of the usual Green's functions in Minkowski spacetime. Otherwise, one can calculate the same correlation functions at an arbitrary temperature using the formalism of thermal field theory, and then take the limit $T\rightarrow T_U$. As will be seen below, the two different methods will give different values for the entropy. 
\\\\
To begin with, we will consider the calculation of correlation functions for the entropy at Unruh temperature $T=T_U$. We will use the method described in \cite {Smolkin:2014hba}. In this paper, the authors present a method for calculating correlation functions in spacetimes with conical defects, including the Rindler spacetime. It was shown that N-point functions in a spacetime with a conical defect can be expressed in terms of $(N+1)$-point functions in an ordinary flat Minkowski spacetime without defects with the additional operator of a modular Hamiltonian, which is essentially a boost operator
\begin{eqnarray}
    \lim_{\nu \to 1} \frac{\partial}{\partial \nu} \langle \mathcal{O}_1(\text{x}_1)... \mathcal{O}_N(\text{x}_N) \rangle_{\nu} = - \langle \mathcal{O}_1(\text{x}_1)... \mathcal{O}_N(\text{x}_N) K_0 \rangle_{conn.},
    \label{technique}
\end{eqnarray}
where $\nu$ is the so called angular deficit, defined as a ratio $\nu = T/T_U$ and $K_0$ is a modular Hamiltonian defined as
\begin{eqnarray}
    K_0=-2 \pi \iint d\text{x} d\text{y} \int_0^{\infty} d\text{z} \,\text{z} \, T_{00}(0,\text{y}) = 2 \pi H_R.
\end{eqnarray}
Note that integration along the z coordinate goes from 0 to $\infty$, which indicates that we are considering only the Right Rindler wedge, which is expressed by the index of the Hamiltonian. Now using formula (\ref{technique}), we can write an equation for the derivative of the average energy-momentum tensor value with respect to the angular deficit of the conical defect
\begin{eqnarray}
    \lim_{\nu \to 1} \frac{\partial}{\partial \nu} \langle \hat{T}_{ij} \rangle = - \langle \hat{T}_{ij} \hat{K}_0 \rangle.
\end{eqnarray}
Since $\nu = 2 \pi T/a$ we can interchange limits as $\lim_{\nu \to 1} \frac{\partial}{\partial \nu} \rightleftarrows \frac{a}{2 \pi} \lim_{T \to T_U} \frac{\partial}{\partial T}$, therefore due to (\ref{sloc}) for the local entropy we have
\begin{eqnarray}
    s_{loc} = \lim_{T \to T_U} \frac{\partial p}{\partial T} = \frac{2 \pi}{a} \lim_{\nu \to 1} \frac{\partial \langle \hat{T}_{ii} \rangle}{\partial \nu} =- \frac{2 \pi}{a} \langle \hat{T}_{ii} \, \hat{K}_0 \rangle.
\end{eqnarray}
It is easy to show that \cite{Smolkin:2014hba}
\begin{eqnarray}
    \langle \hat{T}_{ii} \hat{K}_0 \rangle = C_T \, \frac{\pi^2}{120} \frac{1}{\xi^4}.
\end{eqnarray}
Thus, substituting this expression into the expression for local entropy above and using the rules (\ref{rule}) and (\ref{intrule}), we eventually obtain
\begin{eqnarray}
    s = -C_T \, \frac{\pi^3}{60} \int_{l_c}^{\infty} \frac{d \xi}{\xi^3} = C_T \, \frac{\pi^3}{120 \, l^2_c}.
\end{eqnarray}
Consequently, we see that the entanglement entropy of the causal horizon can be expressed as a function of the central charge $C_T$ of the considered theory.
\\\\
Let us now consider viscosity and return to formulas (\ref{etakubo}) and (\ref{etakubo1}). Since the value of viscosity can also be expressed as a function of the central charge of the considered theory, we must calculate the integrals of expression in brackets in (\ref{G1}). Thus, the invariant expression for viscosity in an arbitrary theory will be in the form
\begin{eqnarray}
    \eta = C_T \, \frac{\pi^2}{480 l^2_c}.
\end{eqnarray}

\subsection{KSS bound fulfillment}

Eventually, our analysis gives us the celebrated ratio
\begin{eqnarray}
    \frac{\eta}{s} = \frac{1}{4 \pi}.
\end{eqnarray}
Despite the negative expressions for the viscosity and entropy values, their ratio leads to the correct value of the lower bound of the ratio (\ref{KSS}). Since both quantities are components of the two-point correlators of the energy-momentum tensor, both expressions contain the constant of the central charge of the theory
\begin{eqnarray}
    C_T^{(RSA)} = - \frac{14}{\pi^4},
\end{eqnarray}
which is negative but reduces in the KSS ratio. Thus, for RSA theory, despite all the problems, the holographic limit (\ref{KSS}) has been formally fulfilled, at least if we prefer to calculate the entropy at $T=T_U$.

\section{Entropy as a three-point correlation function: positive entropy and KSS bound violation}
\label{entropy2}

In this section, we will take a detailed look at calculating the energy-momentum tensor of an accelerated system at an arbitrary temperature. The zero component of the obtained energy-momentum tensor will correspond to the energy density of the system. Then using formula (\ref{sloc}) and then taking the limit of $T\rightarrow T_U$, we obtain the entropy of the considered system. The energy-momentum tensor value of thermal vacuum excitations will be obtained by calculating quantum corrections related to the acceleration of the reference frame. A detailed method for calculating such corrections for acceleration and vorticity is described in \cite{Buzzegoli:2020fjm}. This formalism has already been used many times and has led to key results in \cite{Prokhorov:2019yft, Prokhorov:2021bbv, Prokhorov:2022udo}.
\\\\
Let us briefly introduce the method of calculation. We will consider thermal vacuum excitations as a relativistic fluid in local thermodynamic equilibrium. In the zero approximation, the energy-momentum tensor of a relativistic fluid has the form
\begin{eqnarray}
\label{0-order}
    T^{\mu\nu}= (\rho + p)u^{\mu} u^{\nu} - p g^{\mu\nu}.
\end{eqnarray}
In quantum statistical mechanics, the above expression corresponds to the average value of the quantum tensor energy-momentum operator based on local quantum fields with the Zubarev density operator \cite{Becattini:2015nva}
\begin{eqnarray}
    \hat{\rho} = \frac{1}{Z}\,\exp \{ -\beta_{\mu}(\text{x}) \hat{P}^{\mu} + \frac{1}{2} \varpi_{\mu\nu} \hat{J}^{\mu\nu}_\text{x} \},
    \label{zubarev}
\end{eqnarray}
where $\hat{P}^{\mu}$ is the 4-momentum operator, $\hat{J}^{\mu\nu}_\text{x}$ are the Lorentz transformation generators, shifted by $\text{x}_{\mu}$, and $\beta_{\mu}$ is the four-temperature vector such that the temperature and 4-velocity of the fluid can be defined as
\begin{eqnarray}
    T= \frac{1}{\sqrt{\beta^2}}, \quad
    u_{\mu}(\text{x}) = \frac{\beta_{\mu}(\text{x})}{\sqrt{\beta^2}}
\end{eqnarray}
The generators $\hat{J}^{\mu\nu}_\text{x}$ of the Lorentz transformations shifted to the point $\text{x}$:
\begin{eqnarray}
    \hat{J}^{\mu\nu}_\text{x} = \int d\Sigma_\lambda \left[ (\text{y}^\mu - \text{x}^{\mu}) \, \hat{T}^{\lambda\nu}(\text{y}) - (\text{y}^\nu - \text{x}^{\nu}) \, \hat{T}^{\lambda\mu}(\text{y}) \right],
    \label{SETdefinition}
\end{eqnarray}
where $d \Sigma_{\lambda}$ denotes an element of an arbitrary three-dimensional space-like hypersurface. Operator $\hat{J}^{\mu\nu}$ can be decomposed into boost $\hat{K}^{\mu}$ and orbital angular momentum $\hat{J}^{\mu}$
\begin{eqnarray}
    \hat{J}^{\mu\nu} = u^{\mu} \hat{K}^{\nu} - u^{\nu} \hat{K}^{\mu} - \varepsilon^{\mu\nu\alpha\beta} u_{\alpha} \hat{J}_{\beta}
    \label{Jmunu}
\end{eqnarray}
The quantum corrections associated with the thermal vorticity tensor can be calculated using perturbation theory and finite temperature field theory. The average of the local operator $\hat{O}(\text{x})$ is determined via the perturbative series
\begin{eqnarray}
    \langle \hat{O}(\text{x}) \rangle 
= \langle \hat{O}(0) \rangle_{\beta} 
+ \sum_{N=1}^\infty \frac{\varpi^N}{2^N |\beta|^N N!} 
\int_0^{|\beta|} d\tau_1 \ldots d\tau_N \,
\langle T_\tau \hat{J}^{-i\tau_1 u} \ldots \hat{J}^{-i\tau_N u} \hat{O}(0) \rangle_{\beta,c} \,\,.
\label{series}
\end{eqnarray}
The postscript $\langle ..\rangle_{\beta, c}$ means the connected part of the average of the grand canonical ensemble, $\hat{T}_{\tau}$ means ordering by imaginary time $\tau =it$, the generators are shifted by the vector $-i \tau_n u^{\mu}$ and each of the operators $\hat{J}$ is contracted with one of the thermal vorticity tensors: $\varpi_{\mu\nu}\hat{J}^{\mu\nu}$. The tensor $\varpi_{\mu\nu}$ is called thermal vorticity 
\begin{eqnarray}
    \varpi_{\mu\nu} = - \frac{1}{2} (\partial_{\mu} \beta_{\nu} - \partial_{\nu} \beta_{\mu}).
    \label{varpi}
\end{eqnarray}
It can also be decomposed into familiar acceleration and vorticity. 
\begin{eqnarray}
    \varpi_{\mu\nu} = \varepsilon_{\mu\nu\alpha\beta}\, w^{\alpha} u^{\beta} + \alpha_{\mu} u_{\nu} - \alpha_{\nu} u_{\mu}.
\end{eqnarray}
Therefore, from (\ref{Jmunu}) and (\ref{varpi}) it follows that scalar products with vorticity tensor in (\ref{zubarev}) and (\ref{series}) decompose into terms with boost and angular momentum
\begin{eqnarray}
    \varpi_{\mu \nu} \hat{J}^{\mu \nu}_\text{x} =-2\alpha_{\mu} \hat{K}^{\mu}_\text{x} -2w_{\mu} \hat{J}^{\mu}_\text{x}.
\end{eqnarray}
Further, we will consider uniformly only accelerated media without vorticity, therefore (\ref{zubarev}) transforms to the density operator of the form
\begin{eqnarray}
    \hat{\rho}= \frac{1}{Z} \, \exp \{ -\beta_{\mu} \hat{P}^{\mu} - \alpha_{\mu} \hat{K}^{\mu}_\text{x} \},
\end{eqnarray}
and the perturbation theory in (\ref{series}), takes the form of the series in acceleration
\begin{eqnarray}
    \langle \hat{O}(\text{x}) \rangle = \langle \hat{O}(0) \rangle_{\beta} + \sum_{N=1}^\infty \frac{(-1)^N a^N}{N!} \int_0^{|\beta|} d\tau_1 \ldots d\tau_N \, \langle T_\tau \hat{K}^{-i\tau_1 u} \ldots \hat{K}^{-i\tau_N u} \hat{O}(0) \rangle_{\beta,c} \,\,.
    \label{acc_series}
\end{eqnarray}
Now, using the perturbative series of the statistical density operator (\ref{acc_series}) and calculating corrections to the zero order of the energy-momentum tensor (\ref{0-order}), in general we can obtain the energy-momentum tensor of the vacuum of the theory with nonzero acceleration
\begin{eqnarray}
    \langle \hat{T}_{\mu\nu} \rangle = \Big( A_1 + A_2\, |a|^2 + A_3 \, |a|^4 \Big) \Big( u_{\mu} u_{\nu} - \frac{\Delta_{\mu\nu}}{3}\Big),
    \label{EMT_series}
\end{eqnarray}
where $\Delta_{\mu\nu} = g_{\mu\nu} -u_{\mu}u_{\nu}$. The values of the quantum corrections will be contained in the coefficients $A_2$ and $A_3$. In order to eventually calculate the entropy, it will be necessary to take derivative with respect to the temperature according to (\ref{sloc}). Based on dimensional analysis, the coefficient $A_3$ does not depend on temperature, which means that to calculate entropy, it will be enough for us to know only the coefficients $A_1$ and $A_2$. 

\subsection{Some exact calculation details}

Now, let us look at the calculation of coefficients in energy-momentum tensor in more detail. The first coefficient in (\ref{EMT_series}) according to (\ref{acc_series}) is just
\begin{eqnarray}
    A_1 = \langle \hat{T}^{00} (0) \rangle_{\beta}.
\end{eqnarray}
In general case the thermal averaged energy-momentum tensor for RSA theory will be in the form
\begin{eqnarray}
    \langle \hat{T}^{\mu\nu} \rangle_{\beta} = -Tr \Big[ \mathcal{D}^{\mu\nu\tau\theta}_{\bar{\psi} \psi} \langle \Psi_{\theta} \bar{\Psi}_{\tau}  \rangle_{\beta} + \mathcal{D}^{\mu\nu\tau}_{  \bar{\psi} \lambda } \langle \lambda \bar{\Psi}_{\tau}  \rangle_{\beta} + \mathcal{D}^{\mu\nu\tau}_{\bar{\lambda} \psi} \langle \Psi_{\tau} \bar{\lambda}  \rangle_{\beta} \Big],
\end{eqnarray}
where the minus sign comes from fields permutations in propagators. Formally, the expressions for the differential operators will remain the same as in (\ref{diffoperators}). However, since we will work now in Euclidean spacetime, it will be necessary to replace $\partial_{\mu} \to \tilde{\partial}_{\mu}$, $\gamma_{\mu} \to \tilde{\gamma}_{\mu}$, and $g_{\mu \nu} \to \delta_{\mu \nu}$ and also use expressions for thermal propagators instead of ordinary ones according to the rules (\ref{fourier_propagators}), (\ref{thermalpropagators}) and (\ref{notation}). 
\begin{eqnarray}
    \langle \hat{T}^{\mu \nu} \rangle_{\beta} = - \int \frac{d^3p}{(2 \pi)^3} \, T \, \sum_{ \{ p_n \} } \lim_{m \to \infty} \lim_{X_1,X_2 \to X} Tr \Bigg[ \varepsilon^{\tau\theta\nu\eta}  \Big( \frac{1}{2} \gamma_5 \tilde{\gamma}^{\mu} \tilde{\partial}^{X_2}_{\eta} - \frac{1}{8} \gamma_5 \tilde{\gamma}_{\eta} [\tilde{\gamma}^{\xi},\tilde{\gamma}^{\mu}] (\tilde{\partial}^{X_1}_{\xi} + \tilde{\partial}^{X_2}_{\xi}) \Big) \times \nonumber \\
    \frac{i }{P^2} e^{iP(X_1-X_2)} \Big(  \Tilde{\gamma}_{\tau} \slashed P \Tilde{\gamma}_{\theta} + 2 \, \Big( \frac{1}{m^2}  - \frac{2}{P^2} \Big) P_{\theta} P_{\tau} \slashed P \Big) + 2 \Big( \frac{im}{2} \Tilde{\gamma}^{\mu} \delta^{\nu \tau} \Big) \, e^{iP(X_1-X_2)} \frac{P_{\tau} \slashed P}{m \, P^2} \Bigg] + (\mu \longleftrightarrow \nu).
\end{eqnarray}
Taking the trace and substituting the values of the indices $\mu=\nu=0$, we obtain the expression for $A_1$
\begin{eqnarray}
    A_1= \int \frac{d^3p}{(2 \pi)^3} \, T \sum_{ \{ p_n \} } \frac{8p_n^2-4E^2}{p_n^2+E^2},
    \label{A1}
\end{eqnarray}
where the energy is equal to full momenta $E= \sqrt{p_1^2+p_2^2+p_3^2} =|\textbf{p}|$ since we are working in a massless case. Summation over Matsubara frequencies can be represented integrally as \cite{Buzzegoli:2020fjm}
\begin{eqnarray}
    T \sum_{ \{ p_n \} } f(p_n) = \int_{-\infty}^{\infty} \frac{d\text{z}}{2 \pi} f(\text{z}) - \int_{-\infty-i \varepsilon}^{\infty-i \varepsilon} \frac{d\text{z}}{2 \pi} \Big( f(\text{z})+f(-\text{z}) \Big) n_F(i\text{z}),
\end{eqnarray}
where $n_F$ is the Fermi-Dirac distribution function:
\begin{eqnarray}
    n_F(E) = \frac{1}{e^{|\beta| E}+1}.
\end{eqnarray} 
Then we apply this summation formula to expression (\ref{A1}), discard the first divergent integral, and then take the second integral as the sum of the residues in the lower half-plane $\text{z}=-ip$. As a result, we get the expression
\begin{eqnarray}
    A_1= \int \frac{d^3p}{(2 \pi)^3} \frac{12p}{e^{p/T}+1} = \int^{\infty}_0 dp \, \frac{6p^3}{\pi^2 (e^{p/T}+1)} =\frac{7 \pi^2 T^4}{20}.
\end{eqnarray}
Now, in order to calculate the term $A_2$ we have to expand (\ref{series}) up to a second term, thus
\begin{eqnarray}
    A_2 = \frac{1}{2} \int_0^{|\beta|} d \tau_1 d \tau_2 \langle \hat{K}_{-i\tau_1}^3 \, \hat{K}_{-i\tau_2}^3 \, \hat{T}^{00}(0) \rangle_{\beta}.
    \label{a2}
\end{eqnarray}
For further convenience, we can define the boost generators as an integral of energy-momentum tensor on a timeslice as
\begin{eqnarray}
    \hat{K}^i = \int d^3\text{x} \, \hat{T}_{00} \, \text{x}^i,
\end{eqnarray}
and rewrite (\ref{a2}) in a more general way as a three-point correlator of the energy-momentum tensors, as was made in \cite{Buzzegoli:2017cqy}
\begin{eqnarray}
C^{\alpha\beta|\rho\sigma|\mu\nu|ij}= \frac{1}{|\beta|^2} \int_0^{|\beta|} d \tau_1 d \tau_2 \int d^3 \text{x} \, d^3 \text{y} \, \langle T_{\tau} \{ \hat{T}^{\alpha\beta}(\tau_1, \textbf{x})\,\hat{T}^{\rho\sigma}(\tau_2, \textbf{y})\,\hat{T}^{\mu\nu}(0) \} \rangle_{\beta} \, \text{x}^i \text{y}^j.
\label{Cform}
\end{eqnarray}
Using the notation above, the expression for $A_2$ will be
\begin{eqnarray}
    \label{A2_ind}
    A_2 = \frac{1}{2} C^{00|00|00|33}.
\end{eqnarray}
The essential part of the calculation lies in the calculation of the T-ordered correlator in (\ref{Cform}). We will decompose the energy-momentum tensor as was done in (\ref{SETIK}). The three-point function will have the form (it is assumed that thermal correlators $\langle .. \rangle_{\beta}$ are also considered here)
\begin{eqnarray}
\label{terms3corr}
    \langle \hat{T}^{\alpha\beta} \,\hat{T}^{\rho\sigma} \,\hat{T}^{\mu\nu}  \rangle =  \langle \hat{T}^{\alpha\beta}_{\bar{\psi} \psi} \,\hat{T}^{\rho\sigma}_{\bar{\psi} \psi} \,\hat{T}^{\mu\nu}_{\bar{\psi} \psi}  \rangle + 
    \langle \hat{T}^{\alpha\beta}_{\bar{\psi} \psi} \,\hat{T}^{\rho\sigma}_{\bar{\psi} \psi} \,\hat{T}^{\mu\nu}_{\bar{\psi} \lambda}  \rangle + 
    \langle \hat{T}^{\alpha\beta}_{\bar{\psi} \psi} \,\hat{T}^{\rho\sigma}_{\bar{\psi} \psi} \,\hat{T}^{\mu\nu}_{\bar{\lambda} \psi}  \rangle + 
    \langle \hat{T}^{\alpha\beta}_{\bar{\psi} \psi} \,\hat{T}^{\rho\sigma}_{\bar{\psi} \lambda} \,\hat{T}^{\mu\nu}_{\bar{\psi} \psi}  \rangle + 
    \langle \hat{T}^{\alpha\beta}_{\bar{\psi} \psi} \,\hat{T}^{\rho\sigma}_{\bar{\psi} \lambda} \,\hat{T}^{\mu\nu}_{\bar{\psi} \lambda}  \rangle + \\ \nonumber 
    +\langle \hat{T}^{\alpha\beta}_{\bar{\psi} \psi} \,\hat{T}^{\rho\sigma}_{\bar{\psi} \lambda} \,\hat{T}^{\mu\nu}_{\bar{\lambda} \psi}  \rangle +\langle \hat{T}^{\alpha\beta}_{\bar{\psi} \psi} \,\hat{T}^{\rho\sigma}_{\bar{\lambda} \psi} \,\hat{T}^{\mu\nu}_{\bar{\psi} \psi}  \rangle+
    \langle \hat{T}^{\alpha\beta}_{\bar{\psi} \psi} \,\hat{T}^{\rho\sigma}_{\bar{\lambda} \psi} \,\hat{T}^{\mu\nu}_{\bar{\psi} \lambda}  \rangle+
    \langle \hat{T}^{\alpha\beta}_{\bar{\psi} \psi} \,\hat{T}^{\rho\sigma}_{\bar{\lambda} \psi} \,\hat{T}^{\mu\nu}_{\bar{\lambda} \psi}  \rangle + \langle \hat{T}^{\alpha\beta}_{\bar{\psi} \lambda}\,\hat{T}^{\rho\sigma}_{\bar{\psi} \psi} \,\hat{T}^{\mu\nu}_{\bar{\psi} \psi}  \rangle + 
    \\ \nonumber
    +\langle \hat{T}^{\alpha\beta}_{\bar{\psi} \lambda}\,\hat{T}^{\rho\sigma}_{\bar{\psi} \psi} \,\hat{T}^{\mu\nu}_{\bar{\psi} \lambda}  \rangle + \langle \hat{T}^{\alpha\beta}_{\bar{\psi} \lambda}\,\hat{T}^{\rho\sigma}_{\bar{\psi} \psi} \,\hat{T}^{\mu\nu}_{\bar{\lambda} \psi}  \rangle + \langle \hat{T}^{\alpha\beta}_{\bar{\psi} \lambda}\,\hat{T}^{\rho\sigma}_{\bar{\psi} \lambda} \,\hat{T}^{\mu\nu}_{\bar{\psi} \psi}  \rangle + \langle \hat{T}^{\alpha\beta}_{\bar{\psi} \lambda} \,\hat{T}^{\rho\sigma}_{\bar{\psi} \lambda} \,\hat{T}^{\mu\nu}_{\bar{\psi} \lambda}  \rangle + \langle \hat{T}^{\alpha\beta}_{\bar{\psi} \lambda} \,\hat{T}^{\rho\sigma}_{\bar{\lambda} \psi} \,\hat{T}^{\mu\nu}_{\bar{\psi} \psi}  \rangle + 
    \\ \nonumber
    +\langle \hat{T}^{\alpha\beta}_{\bar{\lambda} \psi} \,\hat{T}^{\rho\sigma}_{\bar{\psi} \psi} \,\hat{T}^{\mu\nu}_{\bar{\psi} \psi}  \rangle + \langle \hat{T}^{\alpha\beta}_{\bar{\lambda} \psi} \,\hat{T}^{\rho\sigma}_{\bar{\psi} \psi} \,\hat{T}^{\mu\nu}_{\bar{\psi} \lambda}  \rangle + \langle \hat{T}^{\alpha\beta}_{\bar{\lambda} \psi} \,\hat{T}^{\rho\sigma}_{\bar{\psi} \psi} \,\hat{T}^{\mu\nu}_{\bar{\lambda} \psi}  \rangle +\langle \hat{T}^{\alpha\beta}_{\bar{\lambda}\psi}\,\hat{T}^{\rho\sigma}_{\bar{\psi} \lambda} \,\hat{T}^{\mu\nu}_{\bar{\psi} \psi}  \rangle + \\ \nonumber
    +\langle \hat{T}^{\alpha\beta}_{\bar{\lambda} \psi} \,\hat{T}^{\rho\sigma}_{\bar{\lambda} \psi} \,\hat{T}^{\mu\nu}_{\bar{\psi} \psi}  \rangle + \langle \hat{T}^{\alpha\beta}_{\bar{\lambda} \psi} \,\hat{T}^{\rho\sigma}_{\bar{\lambda} \psi} \,\hat{T}^{\mu\nu}_{\bar{\lambda} \psi}  \rangle.
\end{eqnarray}
We write down all the correlators that are non-zero along with conditions $\langle \bar{\lambda} \lambda \rangle=0$ and in the limit $m \to \infty$. Overall, there will be 21 non-zero terms.
\\\\
Similarly as in Section \ref{RSAVISCOSITY}, we will consider in detail the algorithm of calculation only of the first term in (\ref{terms3corr}). Other terms can be calculated in the same way. Using formulas (\ref{pointsplit}) for the energy-momentum tensor point splitting procedure and the Wick formula, we have the following expression
\begin{eqnarray}
    C_1(X,Y,Z) = \langle \hat{T}^{\alpha\beta}_{\bar{\psi} \psi}(X) \,\hat{T}^{\rho\sigma}_{\bar{\psi} \psi} (Y)\,\hat{T}^{\mu\nu}_{\bar{\psi} \psi} (Z) \rangle_{\beta} = \quad \quad\quad\quad\quad\quad\quad\quad\quad\quad\quad\quad\quad\quad\quad\quad\quad\quad\quad\quad\quad\quad\quad\quad\quad\quad\quad\quad\quad \\ \nonumber
    =\lim_{X,Y,Z} \mathcal{D}^{\alpha\beta\gamma\theta} (\partial_{X_1}, \partial_{X_2}) \mathcal{D}^{\rho\sigma\delta\tau} (\partial_{Y_1}, \partial_{Y_2})\mathcal{D}^{\mu\nu\eta\xi} (\partial_{Z_1}, \partial_{Z_2}) \, \langle \bar{\Psi}_{\gamma} (X_1) \Psi_{\theta} (X_2) \,  \bar{\Psi}_{\delta} (Y_1) \Psi_{\tau} (Y_2) \, \bar{\Psi}_{\eta} (Z_1) \Psi_{\xi} (Z_2) \rangle_{\beta,c}
    \label{3point}
\end{eqnarray}
The 6-point correlator of fields is decomposed by Wick's theorem as
\begin{eqnarray}
    \langle \bar{\Psi}^{a_1}_{\gamma} (X_1) \Psi_{\theta}^{a_2} (X_2) \,  \bar{\Psi}_{\delta}^{a_3} (Y_1) \Psi_{\tau}^{a_4} (Y_2) \, \bar{\Psi}_{\eta}^{a_5} (Z_1) \Psi_{\xi}^{a_6} (Z_2) \rangle_{\beta,c} = \quad \quad\quad\quad\quad\quad\quad\quad\quad\quad\quad\quad\quad\quad\quad\quad\quad\quad\quad\quad\quad \\ \nonumber
    =- G^{a_6a_1}_{\xi\gamma} (Z_2-X_1)\,G^{a_2a_3}_{\theta\delta} (X_2-Y_1)\,G^{a_4a_5}_{\tau\eta} (Y_2-Z_1) - 
    G^{a_4a_1}_{\tau\gamma} (Y_2-X_1)\,G^{a_2a_5}_{\theta\eta} (X_2-Z_1)\,G^{a_6a_3}_{\xi\delta} (Z_2-Y_1).
\end{eqnarray}
Again, the minus signes come from the Wick theorem and the permutation of spin-3/2 fields in order to make all propagators in expression of the form (\ref{thermalpropagators}). Therefore, the overall expression (\ref{3point}) will be in the form
\begin{eqnarray}
    \langle \hat{T}^{\alpha\beta}_{\bar{\psi} \psi} \,\hat{T}^{\rho\sigma}_{\bar{\psi} \psi} \,\hat{T}^{\mu\nu}_{\bar{\psi} \psi}  \rangle_{\beta} = (-1) \lim_{X,Y,Z} \Bigg( \mathcal{D}^{\alpha\beta\gamma\theta}_{a_1a_2} (\partial_{X_1},\partial_{X_2}) \, G^{a_2a_3}_{\theta\delta} \, \mathcal{D}_{a_3a_4}^{\rho\sigma\delta\tau} (\partial_{Y_1},\partial_{Y_2}) \, G^{a_4a_5}_{\tau\eta}  \, \mathcal{D}_{a_5a_6}^{\mu\nu\eta\xi} (\partial_{Z_1},\partial_{Z_2}) \, G_{\xi\gamma}^{a_6a_1} + \\ \nonumber +\mathcal{D}^{\alpha\beta\gamma\theta}_{a_1a_2} (\partial_{X_1},\partial_{X_2}) \, G^{a_2a_5}_{\theta\eta} \, 
    \mathcal{D}_{a_5a_6}^{\mu\nu\eta\xi} (\partial_{Z_1},\partial_{Z_2}) \, G^{a_6a_3}_{\xi\delta} \, \mathcal{D}_{a_3a_4}^{\rho\sigma\delta\tau} (\partial_{Y_1},\partial_{Y_2}) \, G_{\tau\gamma}^{a_4a_1} \Bigg).
\end{eqnarray}
In this case, unlike the calculation of viscosity, the propagators in the expression will be thermal
\begin{eqnarray}
    G_{\mu_1\mu_2}^{a_1 a_2} (X_1-X_2) = \langle T_{\tau} \Psi_{\mu_1}^{a_1}(X_1) \Bar{\Psi}^{a_2}_{\mu_2}(X_2) \rangle_{\beta}
\end{eqnarray}
In this case, it will be more convenient for us to work in the momentum representation, so we will write the thermal propagators in the Fourier representation \cite{Prokhorov:2021bbv}
\begin{equation}
    \label{fourier_propagators}
    \begin{cases}
        G^{\psi \bar{\psi}}_{\mu \nu} (X_1-X_2) = \langle T_{\tau} \Psi_{\mu}(X_1) \Bar{\Psi}_{\nu}(X_2) \rangle_{\beta} = \sumint  e^{i P (X_1-X_2)} \, G^{\psi \bar{\psi}}_{\mu \nu} (P),\vspace{0.2 cm} \\
        G^{\lambda \bar{\psi}}_{\mu} (X_1-X_2) = \langle T_{\tau} \lambda(X_1) \Bar{\Psi}_{\mu}(X_2) \rangle_{\beta} =  \sumint  e^{i P (X_1-X_2)} \, G^{\lambda \bar{\psi}}_{\mu} (P),\vspace{0.2 cm} \\
        G^{\psi \bar{\lambda}}_{\mu} (X_1-X_2) = \langle T_{\tau} \Psi_{\mu}(X_1) \Bar{\lambda}(X_2) \rangle_{\beta} =  \sumint  e^{i P (X_1-X_2)} \, G^{\psi \bar{\lambda}}_{\mu} (P),
    \end{cases}
\end{equation}
where the propagators in momentum representation have the form
\begin{equation}
    \label{thermalpropagators}
    \begin{cases}
        G^{\psi \bar{\psi}}_{\mu \nu} (P) =  \frac{i}{P^2}
        \Bigg( \Tilde{\gamma}_{\nu} \slashed P \Tilde{\gamma}_{\mu} + 2 \, \Big( \frac{1}{m^2} - \frac{2}{P^2} \Big) P_{\mu} P_{\nu} \slashed P \Bigg),\vspace{0.2 cm} \\
        G^{\lambda \bar{\psi}}_{\mu} (P) =  \frac{P_{\mu} \, \slashed P}{m \, P^2},\vspace{0.2 cm} \\
        G^{\psi \bar{\lambda}}_{\mu} (P) = \frac{-P_{\mu} \, \slashed P}{m \, P^2}.
    \end{cases}
\end{equation}
Since we are now working in the formalism of thermal field theory, it will be necessary to move on to the Euclidean signature of the metric tensor. Therefore, for convenience, the following notation is introduced:
\begin{eqnarray}
X^\mu = (\tau_x, -\textbf{x}), \quad \gamma^\mu = i^{\delta^{0\mu-1}} \tilde{\gamma}^\mu, 
\quad \tilde{\gamma}_5 = \gamma_5 = i\gamma^0 \gamma^1 \gamma^2 \gamma^3, \quad \partial_\mu = i^{\delta^{0\mu}} \tilde{\partial}_\mu, \quad \tilde{\partial}_X^\mu = \left(\frac{\partial}{\partial \tau_\text{x}}, \,\frac{\partial}{\partial \text{x}}\right), \quad \psi_\mu = i^{\delta^{0\mu}} \tilde{\psi}_\mu, \nonumber \\
P^\mu = (p_n, -\textbf{p}), \quad p_n = (2n+1)\pi T, \quad n = (0, \pm 1, \ldots),  \quad \sumint_{\substack{P}} = T \sum_{p_n} \int \frac{d^3p}{(2\pi)^3}, \quad \slashed{P} = P^\mu \tilde{\gamma}_\mu, \quad P^2 = P_\mu P^\mu ,
\label{notation}
\end{eqnarray}
where $p_n$ are the fermionic Matsubara frequencies and Euclidean gamma matrices $\tilde{\gamma}_{\mu}$ satisfy the relation $\{\tilde{\gamma}_{\mu},\tilde{\gamma}_{\nu}\}=2\delta_{\mu\nu}$.
Use operators (\ref{diffoperators}) with replacement $\partial_{\mu} \to \tilde{\partial}_{\mu}$, $\gamma_{\mu} \to \tilde{\gamma}_{\mu}$, and $\eta_{\mu \nu} \to \delta_{\mu \nu}$ according to (\ref{notation}). 
\\\\
Plugging expressions for operators and propagators in (\ref{3point}) we have
\begin{eqnarray}
    C_1(X,Y,Z) = (-1) \sumint_{P} \sumint_{Q} \sumint_{K} e^{iP(X-Y)+iQ(Y-Z)+iK(X-Z)} \Big[ F^{(1)}_1(P,Q,K) + F^{(2)}_1(P,Q,K)\Big].
    \label{bigC}
\end{eqnarray}
We defined functions $F^{(1)}_1$ and $F^{(2)}_1$ as
\begin{eqnarray}
    \label{bigF0}
    F^{(1)}_1 (P,Q,K) = Tr \Big[ \mathcal{D}^{\alpha\beta\gamma\theta}_{a_1a_2} (-iK, iP) \, G^{a_2a_3}_{\theta\delta}(P) \, \mathcal{D}_{a_3a_4}^{\rho\sigma\delta\tau} (-iP,iQ) \, G^{a_4a_5}_{\tau\eta} (Q)  \, \mathcal{D}_{a_5a_6}^{\mu\nu\eta\xi} (-iQ,iK) \, G_{\xi\gamma}^{a_6a_1}(K) \Big], \\ \nonumber
    F^{(2)}_1 (P,Q,K) = Tr \Big[ \mathcal{D}^{\alpha\beta\gamma\theta}_{a_1a_2} (-iK,iP) \, G^{a_2a_5}_{\theta\eta}(K) \, 
    \mathcal{D}_{a_5a_6}^{\mu\nu\eta\xi} (-iQ,iK) \, G^{a_6a_3}_{\xi\delta}(Q) \, \mathcal{D}_{a_3a_4}^{\rho\sigma\delta\tau} (-iP,iQ) \, G_{\tau\gamma}^{a_4a_1}(P) \Big].
\end{eqnarray}
The lower indexes indicate the order of the terms in (\ref{terms3corr}), whereas the higher ones point out order of Wick contraction. Following the procedure from \cite{Becattini:2023ouz}, we describe the expression (\ref{bigC}) in more detail and set the coordinate $Z \to 0$ in order to satisfy (\ref{acc_series})
\begin{eqnarray}
    C_1(X,Y,0) = C_1(\textbf{x},\textbf{y}, \tau_1,\tau_2) = - \int \frac{d^3p}{(2 \pi)^3} \int \frac{d^3q}{(2 \pi)^3} \int \frac{d^3k}{(2 \pi)^3} e^{-i(\textbf{p}+\textbf{k}) \textbf{x}} \, e^{-i(\textbf{q}-\textbf{p}) \textbf{y}} \, S_1(\textbf{p},\textbf{q},\textbf{k}, \tau_1, \tau_2).
    \label{getout}
\end{eqnarray}
We define $S_1(\textbf{p},\textbf{q},\textbf{k}, \tau_1, \tau_2)$ as
\begin{eqnarray}
    S_1(\textbf{p},\textbf{q},\textbf{k}, \tau_1, \tau_2) = T^3 \sum_{ \{ p_n,q_n,k_n\} }  e^{ip_n(\tau_1-\tau_2)} \, e^{iq_n \tau_2} \, e^{ik_n \tau_1}  \Big( F^{(1)}_1(P,Q,K)+F^{(2)}_1(P,Q,K) \Big).
    \label{bigS}
\end{eqnarray}
Note that each spin-3/2 fields propagator (\ref{thermalpropagators}) contains poles of $2^{nd}$ and $4^{th}$ order
\begin{eqnarray}
    G(P) = G_1(P) + G_2(P), \quad G_1 \sim 1/P^2, \quad G_2 \sim 1/P^4.
\end{eqnarray}
Thus, (\ref{bigS}) contain terms with poles of the second and fourth order. The sums over fermionic Matsubara frequencies can be done using formulas from \cite{Prokhorov:2022snx} for higher orders of degrees of the denominator
\begin{equation}
    \begin{cases}
        T \sum_{ \{ p_n\} } \frac{(p_n)^k \, e^{i p_n \tau} }{p_n^2+E^2} = \frac{1}{2E} \sum_{s= \pm 1} (-isE)^k \, e^{\tau s E} \Big[ \theta(-s \tau) - n_F(E) \Big] , \vspace{0.2 cm} \\
        T  \sum_{ \{ p_n\} } \frac{(p_n)^k \, e^{i p_n \tau} }{(p_n^2+E^2)^2} = \sum_{s= \pm 1} e^{\tau s E} 
        \Bigg( \frac{(-isE)^k}{4E^2} \frac{\partial n_F(E)}{\partial E} + \frac{(1-s \tau E)(-isE)^k + isE \frac{\partial (-isE)^k}{\partial E} }{4E^3} \Big[ \theta(-s \tau) - n_F(E) \Big] \Bigg).
    \end{cases}
    \label{sums}
\end{equation}
Note that since $S_1(\textbf{p},\textbf{q},\textbf{k}, \tau_1, \tau_2)$ does not depend on coordinates, we can take advantages of the formula
\begin{eqnarray}
    \int d^3 \text{x} \int d^3 \text{y} \, e^{-i(p+k)\text{x}} e^{-i(p-q)\text{y}} \, \text{x}^i \text{y}^j = -(2 \pi)^6 \frac{\partial^2}{ \partial k_i \partial q_j} \delta^{(3)}(p+k) \, \delta^{(3)}(p-q).
\end{eqnarray}
Therefore, (\ref{Cform}) for the first term in (\ref{terms3corr}) will have the form
\begin{eqnarray}
    C_1^{\alpha\beta|\rho\sigma|\mu\nu|ij}= \frac{1}{|\beta|^2} \int_X \int_Y \, \langle T_{\tau} \{ \hat{T}^{\alpha\beta}(X)\,\hat{T}^{\rho\sigma}(Y)\,\hat{T}^{\mu\nu}(0) \} \rangle_{\beta} \, \text{x}^i \text{y}^j = \\ \nonumber
    =\frac{-1}{|\beta|^2} \iint_0^{|\beta|} d \tau_1 d \tau_2 \int \frac{d^3p}{(2 \pi)^3} \frac{\partial^2}{\partial k_i \partial q_j} S_1(\textbf{p},\textbf{q},\textbf{k}, \tau_1, \tau_2) \Bigg|_{ \textbf{q} = \textbf{p}, \textbf{k} = -\textbf{p}}.
    \label{finalC}
\end{eqnarray}
When we do exactly the same procedure for the sum of all terms in (\ref{terms3corr}) according with (\ref{bigS}), we will get the expression
\begin{eqnarray}
    C^{\alpha\beta|\rho\sigma|\mu\nu|ij}=\frac{-1}{|\beta|^2} \iint_0^{|\beta|} d \tau_1 d \tau_2 \int \frac{d^3p}{(2 \pi)^3} \frac{\partial^2}{\partial k_i \partial q_j} \Bigg[  T^3 \sum_{ \{ p_n,q_n,k_n\} } e^{ip_n(\tau_1-\tau_2)} \, e^{iq_n \tau_2} \, e^{ik_n \tau_1} \times \nonumber \\
    \times \Big( F^{(1)}_1(P,Q,K)+F^{(2)}_1(P,Q,K) + ... + F^{(1)}_{21}(P,Q,K)+F^{(2)}_{21}(P,Q,K) \Big) \Bigg] \Bigg|_{ \textbf{q} = \textbf{p}, \textbf{k} = -\textbf{p}}.
    \label{C_all}
\end{eqnarray}
We can now substitute values of operators (\ref{diffoperators}) and propagators (\ref{thermalpropagators}) into all the functions $F^{(1)}$ and $F^{(2)}$ from the expression above in a similar manner as in (\ref{bigF0}). Then tracing over the spinor indexes and choose Lorentzian indexes to be $\alpha=\beta=\rho=\sigma=\mu=\nu=0$ in order to saturate condition (\ref{A2_ind}). Sequentially performing summation on Matsubara frequencies using rules (\ref{sums}), making momenta differentiations, momenta substitution and all the integrations as in (\ref{C_all}), we finally get the final expression for $A_2$:
\begin{eqnarray}
    A_2 = \frac{1}{2} C^{00|00|00|33} = \frac{1}{12 \pi^2 T^4} \int_0^{\infty} dp  \frac{\, p \, e^{ p/T } }{(1+e^{p/T} )^5} \Bigg( -47 -16 \, p \,T- p^2 + \Big[-47 +48 \, p \,T +11 p^2 \Big] e^{\frac{p}{T}} + \\ \nonumber 
    +\Big[47 +48 \, p \,T -11 p^2 \Big] e^{\frac{2p}{T}} +\Big[47 -16 \, p \,T + p^2 \Big] e^{\frac{3p}{T}} \Bigg) = \frac{T^2}{8}.
    \label{MAINCOEF}
\end{eqnarray}

\subsection{Positive entropy and KSS bound violation}

So we got that in the case of the RSA theory, the energy-momentum tensor of the thermal vacuum state with quantum corrections will have the form
\begin{eqnarray}
    \langle \hat{T}_{\mu\nu} \rangle = \Big( \frac{7 \pi^2 T^4}{20} + \frac{|a|^2 T^2}{8} + O(|a|^4) \Big) \Big( u_{\mu} u_{\nu} - \frac{\Delta_{\mu\nu}}{3}\Big).
    \label{EMTRSA}
\end{eqnarray}
Therefore, the vacuum energy density is
\begin{eqnarray}
    \epsilon = \langle T_{00} \rangle = \Big( \frac{7 \pi^2 T^4}{20} + \frac{|a|^2 T^2}{8} +O(|a|^4)\Big).
    \label{enot}
\end{eqnarray}
An important property of the vacuum energy-momentum tensor is its general covariance, that is, its independence from the reference frame. Since the vacuum energy in the reference frame at rest is zero, it is necessary that condition $\langle \hat{T}_{\mu\nu} \rangle (T=T_U) = 0$ is fulfilled \cite{Becattini:2017ljh}. Note that this condition is also fulfilled for the case of lower spin fields \cite{Prokhorov:2019yft}. Therefore, knowing the coefficients $A_1$, $A_2$ and using the general covariance condition, we can easily restore the coefficient $A_3$. Thus, the whole expression for the vacuum energy of the RSA fields in the accelerated reference frame will have the form
\begin{eqnarray}
    \epsilon^{RSA} = \langle T_{00} \rangle = \Big( \frac{7 \pi^2 T^4}{20} + \frac{|a|^2 T^2}{8} - \frac{17 |a|^4}{320 \pi^2} \Big).
    \label{RSAenergy}
\end{eqnarray}
As we can see, obtained energy-momentum tensor is exactly three times the value for the energy-momentum tensor of the spinor field \cite{Prokhorov:2019yft}
\begin{eqnarray}
    \epsilon^{(s=1/2)}  = \Big( \frac{7 \pi^2 T^4}{60} + \frac{|a|^2 T^2}{24} - \frac{17 |a|^4}{960 \pi^2} \Big) = \frac{1}{3} \epsilon^{RSA}.
\end{eqnarray}
Now, from the equation of the heat balance and the perfect fluid condition $\epsilon = 3 p$, we can obtain the expression for the local volumetric entropy by taking the derivative of (\ref{enot})
\begin{eqnarray}
    s_{loc} (T, a)=\frac{1}{3} \frac{\partial \epsilon}{\partial T} = \frac{7 \pi^2T^3}{15} + \frac{|a|^2T}{12}
\end{eqnarray}
In order to move from the local volumetric entropy to the entropy of the horizon, we make a substitution of the Unruh temperature $T = a/2\pi$ and then take a limit $|a| \to 1/\xi$
\begin{eqnarray}
    s_{loc} (\xi) = \frac{1}{10 \pi \xi^3},
\end{eqnarray}
and then integrate it from the cutoff scale $l_c$ to infinity as was done in \cite{Chirco:2010xx}
\begin{eqnarray}
    s = \int_{\substack{l_c}}^{\infty}  \, s_{loc}(\xi) \, d \xi = \int_{\substack{l_c}}^{\infty}  \, \frac{d \xi}{10 \pi \xi^3}
\end{eqnarray}
Eventually, we have
\begin{eqnarray}
    s = \frac{1}{20 \pi l^2_c}.
\end{eqnarray}
Now, if we compare this with our result for viscosity (\ref{viscosity}), we will find that the KSS ratio is no longer valid and even less than zero, because we are dealing with negative viscosity
\begin{eqnarray}
    \frac{\eta}{s} = - \frac{7}{12 \pi} < 0.
\end{eqnarray}
Despite the fact that the correlators for evaluating entropy can be calculated in two different ways in different regimes, there were no discrepancies for theories with lower spins. In \cite{Lapygin:2025zhn}, entropy was calculated in the limit $T\to T_U$ for theories with spins 0, 1/2, and 1, but the KSS bound equality (\ref{KSS}) was fulfilled. The reason for this discrepancy in the results of calculations in different modes $T=T_U$ and $T\to T_U$ for the case of spin-3/2 fields, apparently, lies in the peculiarities of the Rarita-Schwinger theory on manifolds with a conical singularity. We will discuss this issue in more detail in the section \ref{B}.

\section{Discussion}
\label{discussion}

\subsection{Negative viscosity in RSA theory}
\label{A}

As already noted above, the KSS relation holds in the case of lower spin fields 0, 1/2 and 1 \cite{Lapygin:2025zhn}, while the values of viscosity and entropy in these theories are positive. Despite the fact that in the case of the RSA theory the condition (\ref{KSS}) is formally fulfilled, the calculated shear viscosity takes a negative value (\ref{viscosity}). Taking this result literally, we can say that the disturbances of the Rindler horizon are not damped, but rather increase. From a physical point of view, this may indicate the well-known problems of non-Hermitian and superluminal signals in theory \cite{Velo:1969bt}. It is also interesting to note that the spin-1/2 $\lambda$-fields introduced to eliminate some limitations of the Rarita-Schwinger theory make a negative contribution 
\begin{eqnarray}
    \eta = \eta_{\psi} + \eta_{\lambda} = \frac{31}{240 \pi^2 \, l_c^2} + \frac{-38}{240 \pi^2 \, l_c^2} = \frac{-7}{240 \pi^2 \, l_c^2}
\end{eqnarray}
Thus, failure to meet the positivity condition for the shear viscosity indicates the presence of fundamental consistency problems that accompany the interaction of spin-3/2 fields.

\subsection{Discrepancy between calculation methods and the problem of theories with higher spins}
\label{B}

In this paper, we calculated the entropy of the Rindler horizon in two different ways: the first ensured the fulfillment of condition (\ref{KSS}), but gave a negative value for the entropy. The second method showed that the entropy on the horizon would still be positive, but the result (\ref{KSS}) was not satisfied. An important circumstance is the fact that different ways of calculating entropy in the case of spin-3/2 theory give different results. First, we note that, for lower spin fields, both methods of calculating entropy give the same result, thereby eliminating doubts about the correctness of the methods. The difference in the results of calculations of energy-momentum tensor correlators apparently lies in the features of quantum theories with spin 3/2 on manifolds with a conical singularity (which also includes the Rindler spacetime with horizon). 
\\\\
This property is discussed in sufficient detail in \cite {Fursaev:1996uz}. It is shown that for the higher-spin fields on such manifolds there are appearance of additional “surface" contributions to the heat-kernel in the Schwinger-DeWitt decomposition due to special modes localized at the top of the cone. As was shown, the Rarita–Schwinger operator $\Delta^{(3/2)}$ loses certain local Killing symmetries on the cone, as a result of which additional eigenmodes persist even in the limit of the vanishing deficit angle $\beta \to 2 \pi$. These modes generate a non-vanishing surface term in the first coefficient of the heat-kernel $A_1^{(3/2)}$, which manifests itself as a localized contribution on the singular surface. When the metric is changed, this term creates contact contributions to the correlators of the energy momentum tensors. Thus, for fields with spin 3/2, including in the RSA model, such a localized contribution may have a negative spectral weight. Therefore, through the Kubo relation
\begin{eqnarray}
    \eta = \lim_{\omega \to 0} \frac{1}{\omega} \, Im \, \langle T^{\mu\nu}_{xy} T^{\alpha\beta}_{xy} \rangle(\omega), 
\end{eqnarray}
such a negative contact coefficient may lead to a negative value of the shear viscosity of the Rindler horizon.

\subsection{Spin universality}
\label{C}

The consideration of quantum fields in Rindler spacetime is in some sense similar to the general formulation of the Casimir effect. Since both cases are, in fact, flat spacetimes with boundaries and boundary conditions, it is expected that the expressions for vacuum energy in such systems will be similar \cite {Brown:1986jy}. In work \cite{Stokes:2014pha}, the idea of the spin universality of the Casimir effect is expressed, the authors claim that the vacuum energy of quantum fields with boundaries depends only on the field statistics. Thus, for example, the vacuum energy for fields with spin 1/2 and spin 3/2 Rarita-Schwinger fields with will be the same. Now, if we consider the result that we obtained for the vacuum energy of the fields in RSA theory (\ref{RSAenergy}), we can see that the energy value will exceed the value for Dirac fields by 3 times. However, since in the RSA model two additional degrees of freedom are associated with two additional 1/2-spin $\lambda$-fields, the energy of the 3/2-spin field $\psi{\mu}$ itself coincides with the vacuum energy of the spinor field
\begin{eqnarray}
\label{equality}
    \epsilon^{(s=1/2)} = \epsilon^{(s=3/2)} = \Big( \frac{7 \pi^2 T^4}{60} + \frac{|a|^2 T^2}{24} - \frac{17 |a|^4}{960 \pi^2} \Big).
\end{eqnarray}
That is, we can say that our result (\ref{RSAenergy}) is compatible with the idea expressed in \cite{Stokes:2014pha}.

\subsection{Different energy values for 3/2-spin theory}
\label{D}

It is also of interest to compare the energy-momentum tensor we obtained for a massless field with spin 3/2 with the predictions of other approaches in the literature. It should be noted that existing calculations of the energy-momentum tensor for spin-3/2 fields in an accelerated frame give different conflicting results \cite{Candelas:1978gg, Dowker:1987pk}, and this problem has not yet received a definitive solution. Our formula, obtained within the quantum-statistical approach, differs from both known approaches \cite{Candelas:1978gg, Dowker:1987pk}.
\\\\
Namely, in papers \cite{Candelas:1978gg, Candelas:1978gf, Dowker:1983nt}, the authors provide universal formulas for calculating the vacuum acceleration-dependent $a^4$ contribution to energy, which differ from each other. More specifically, a universal spin-dependent formula was proposed in \cite{Candelas:1978gg, Candelas:1978gf}
\begin{eqnarray}
    \langle T^{\nu}_{\mu} \rangle = - \frac{h(s)}{2 \pi^2 \xi^4} \int_0^{\infty} d \nu \, \frac{\nu \, (\nu^2+s^2)}{e^{2 \pi \nu}-(-1)^{2s}} \, diag \Big(-1,\frac{1}{3},\frac{1}{3},\frac{1}{3} \Big),
\end{eqnarray}
while Dowker writes the general formula \cite{Dowker:1983nt}
\begin{eqnarray}
    E_s=(-1)^{2s} \frac{h(s) \, a^4}{480 \pi^2} \Big( 30s^4-20s^2+1 \Big).
\end{eqnarray}
Despite the fact that these formulas give the same answers for spins 0, 1/2 and 1, they give different answers for the case $s=3/2$. In this context, we are considering the Rarita-Schwinger theory, without additional Adler $\lambda$-fields. As our analysis (\ref{equality}) shows, the correction of the $4^{th}$ order of acceleration will have the form (we use the notation from the expression (\ref{EMT}))
\begin{eqnarray}
    A_3 = - \frac{17}{960 \pi^2},
\end{eqnarray}
while applying the formulas above gives the results
\begin{eqnarray}
    A^{Candelas}_3 = - \frac{97 }{960 \pi^2}, \quad \quad
    A^{Dowker}_3 =  \frac{863}{960 \pi^2}.
\end{eqnarray}
If we now take the same value of the vacuum energy (\ref{equality}) for the Rarita-Schwinger theory, knowing the coefficients $A_3$ written above and based on the principle of covariance $\hat{T}_{\mu\nu}(T=T_U)=0$, we can calculate the coefficients $A_2$ for the Rarita-Schwinger theory. By adding to the result the doubled $A_2$ coefficients for Dirac theory (\ref{equality}), it is possible to obtain results for $A_2$ in RSA in order to compare it with (\ref{MAINCOEF}). Therefore, we have
\begin{eqnarray}
    A^{Candelas}_2 = \frac{11}{24}, \quad \quad
    A^{Dowker}_2 =  -\frac{85}{24}.
\end{eqnarray}
Obviously, obtained results are different and do not coincide with our calculation (\ref{MAINCOEF}). This discrepancy can also be explained by the arguments given in \ref{B}.


\subsection{Bulk-boundary correspondence in Rindler space and ``wandering'' Planck constant}
\label{E}

An analysis of entanglement viscosity allows us to draw a conclusion about the correspondence between the membrane's properties and the thermal radiation above it. As mentioned earlier, an analysis of classical gravitational equations in space with a stretched horizon allows us to interpret the black hole's surface as a membrane with hydrodynamic properties \cite{Parikh:1997ma}. The viscosity of such a membrane depends only on the black hole's area, and its ratio to entropy (described by the well-known Bekenstein-Hawking formula) saturates the bound (\ref{KSS}).
\\\\
In contrast, we consider not the membrane itself (in our case, for the Rindler horizon), but the quantum thermal radiation existing above it - Unruh radiation. At first glance, the correspondence between the membrane's properties and the thermal radiation is not at all obvious. For example, the radiation is characterized by entanglement entropy. Although this entropy is also proportional to the horizon's area \cite{Srednicki:1993im, Bombelli:1986rw}, unlike the entropy in the Bekenstein-Hawking formula, its specific value depends on the type of particles under consideration. Thus, we can speak of the so-called ``species problem'' \cite{Chirco:2010xx}. Similarly, for entanglement viscosity, it differs for different particle types, as can be seen from (\ref{viscosity}) and formulas from \cite{Lapygin:2025zhn, Prokhorov:2026swu, Chirco:2010xx}.
\\\\
The key observation is that, despite this, the ratio of entanglement viscosity to entanglement entropy is universal and the same as the ratio of viscosity to entropy of the membrane itself. Moreover, this ratio is independent of the regularization parameter ($l_c$ in our case), which indicates that it has physical meaning. Thus, we see a correspondence between the properties of the membrane and the medium above it.
\\\\
Note that this correspondence is realized in a rather nontrivial manner. If we restore the dimensional constants, the bound (\ref{KSS}) takes the form $ \eta/s = \hbar/(4 \pi k_B) $, where $ \hbar $ indicates the quantum origin of the bound. However, if for the membrane the contribution $ \hbar $ comes from the black hole entropy $ S_{BH} \sim \mathcal{O}(\hbar^{-1}) $, and the viscosity itself is purely classical $ \eta \sim  \mathcal{O}(\hbar^0) $, then in our case of the entanglement viscosity $ \eta \sim \mathcal{O}(\hbar) l_c^{-2}  $ and the entropy density $ s \sim \mathcal{O}(\hbar^{0}) l_c^{-2}  $, and the parameter $l_c$, generally speaking, is not necessarily small and simply drops out of the relation. This allows us to speak of a kind of ``wandering'' Planck constant in the context of formula (\ref{KSS}) - Planck's constant comes from entropy (for a membrane) or from the viscosity itself (for entanglement viscosity).
\\\\
The physical interpretation of KSS ratio and wandering Planck constant may also be considered  from the point of view of {\it rotational} degrees of freedom. Indeed, the KSS ratio estimate  
\begin{eqnarray}
\frac{\eta}{s} \sim \frac{\rho v l}{n}  = m v l \geq \hbar
\label{interp}
\end{eqnarray}
This can be correlated with the action of the mass particle $m$ propagating along a free path d$l$ with the average velocity $v$ (provided the same components of the free path and velocity are averaged), which is bounded from below by the quantum action $\hbar$. At the same time, if orthogonal components contribute, it may be considered as an average angular momentum of the vortex of size $l$ formed by particle moving with the velocity $v$, which is bounded by the {\it angular momentum} quantum $\hbar$.  
\\\\
The latter interpretation may be compared to the famous Kolmogorov cascade picture, when the energy is transmitted from the larger to smaller vortices, while the smallest ones are responsible for the viscosity effects and dissipation. The KSS bounds therefore relate the 
size of the smallest vertices to the lowest possible viscosity. It is interesting that (\ref{interp}) may also be considered as a bound for velocity circulation in the superfluid. The related interpretation is as follows: the decrease of viscosity below KSS bound may be achieved by transition to the superfluid phase (c.f. \cite{Rupak:2007vp}).
One should also mention that pionic superfluidity \cite{Teryaev:2017wlm,Teryaev:2022pcz} (with the velocity circulation quantum containing chemical potential rather than particle mass) leads to chiral vortical effects and baryon polarization in heavy-ion collisions. 
\\\\
This rotational interpretation of KSS bound corresponds to the "traditional" medium model, rather than the one related to entanglement, which is the subject of this paper. At the same time, the mentioned "wandering" effect may be explained by relation to the famous anyons \cite{Chen:1989xs} concept. Indeed, there is no quantization of angular momentum at two dimensions, and, accordingly, no Planck constant in the numerator of KSS ratio.

\section{Conclusion}
We calculated the entanglement shear viscosity for Unruh radiation in the case of Rarita-Schwinger-Adler model with massless spin-3/2 field. In contrast to the case with lower spins (0, 1/2, 1), the shear viscosity turns out to be negative, which is a direct consequence of the negativity of the conformal central charge.
\\\\
We also calculated the entropy density of the corresponding radiation, considering it as a response to a small angular deficit and applying the method of expansion in powers of the modular Hamiltonian. Within this approach, the entropy density is also negative, similar to the viscosity. But the ratio of the shear viscosity to the entropy density remains positive and saturates the KSS bound, similar to lower spins. Thus, albeit in a non-trivial way with negative viscosity and entropy, we demonstrate the universality of the KSS ratio for entanglement viscosity, which turns out to be valid for higher spins.
\\\\
However, the key question is the origin and meaning of the resulting negative viscosity and entropy. To better understand this, we calculated the entropy density using a different method, based on the Zubarev density operator. As a result, we show that within the framework of this method the problem of negativity, at least for entropy, is overcome and it turns out to be positive. 
This situation seems somewhat unexpected, since for lower spins both methods gave the same result. 
\\\\
Moreover, under the assumption that entropy is additive for different degrees of freedom in the RSA model, we predict within the Zubarev operator method that the entropy for a massless spin-3/2 field coincides with that of a massless spin-1/2 field. We demonstrated parallels between this signature of possible spin universality and the predicted possible spin universality in the Casimir effect.
\\\\
Furthermore, we compared the obtained within the quantum-statistical method energy-momentum tensor for an accelerated medium with known calculations in \cite{Candelas:1978gg} and \cite{Dowker:1987pk}. Assuming normalization to the Minkowski vacuum, we predict that, within the quantum-statistical approach, the energy-momentum tensor for spin 3/2 field is to be equal to the same tensor for the Dirac field, which differs both from  \cite{Candelas:1978gg} and \cite{Dowker:1987pk}. 
\\\\
A natural question arises as to why different methods, which coincide for lower spins, give different results for the energy-momentum tensor and the entropy of spin 3/2, and whether it is possible to say which of the methods is correct. 
Overall, these questions remain open and rather point to difficulties in describing spin 3/2 fields in spaces with a conical singularity. We demonstrate, that it can be compared with the known discrepancy between the off-shell and on-shell methods for calculating the Schwinger-DeWitt coefficients, observed in the work \cite{Fursaev:1996uz}, and to known problems when considering spin-3/2 fields in an external electromagnetic field background \cite{Adler:2017shl}. 
\\\\
At the same time, we explicitly demonstrated that the Rarita-Schwinger-Adler theory in flat space possesses many features of a conformal field theory, unlike conventional Rarita-Schwinger fields. In particular, the energy-momentum tensor trace is zero, and the 2-point one-loop correlator (as well as the 3-point one considered in \cite{Prokhorov:2022rna}) has a form that satisfies the predictions of conformal symmetry. This suggests that the RSA theory provides a conformally-symmetric formulation of the massless spin-3/2 field, but with a negative conformal central charge.

\bibliography{lit}

\end{document}